\documentclass[
reprint,
amsmath,amssymb,
aps, physrev,
nofootinbib
]{revtex4-2}

\usepackage{graphicx}
\usepackage{dcolumn}
\usepackage{bm}
\usepackage{float}
\usepackage{amsmath}
\usepackage{hyperref}
\usepackage{color}
\usepackage{aas_macros}

\begin{document}

\title{Probing Time-Dependent Physics with Phase-Folding CMB Maps}

\author{Yilun Guan}
\email{yilun.guan@utoronto.ca}
\affiliation{Dunlap Institute for Astronomy and Astrophysics, University of Toronto, 50 St. George St., Toronto, ON M5S 3H4, Canada}

\date{\today}

\begin{abstract}
Time-resolved observations of the Cosmic Microwave Background (CMB) offer a powerful probe of time-dependent cosmological signals, such as a stochastic gravitational wave background passing through Earth, which imprints a time-varying deflection on the CMB, and time-dependent cosmic birefringence, which induces an oscillating polarization rotation. However, analyses based on time-division CMB maps are fundamentally limited by the mapmaking cadence, restricting sensitivity to frequencies below $\sim 10^{-5}$\,Hz. In this paper we develop a phase-folding mapmaking framework to enable targeted searches of such periodic cosmological signals with frequencies up to the detector sampling rate of $\mathcal{O}(100)$\,Hz. We demonstrate the power of this framework with two cosmological applications: (1) constraining a stochastic gravitational wave background via its time-dependent lensing signature, and (2) the search for an oscillating polarization rotation from axion-like particles. We show that this technique transforms CMB experiments into broadband probes of the oscillating sky, extending their constraining power from the microhertz regime up to $\mathcal{O}(100)$\,Hz—an expansion of over seven orders of magnitude in frequency. This provides a new observational window, complementary to other approaches by probing under-explored frequency ranges for gravitational waves and axion-like particles.
\end{abstract}

\maketitle

\section{Introduction}
\label{sec:intro}
The Cosmic Microwave Background (CMB) is typically treated as a static snapshot of the early universe. However, the time-ordered data from which CMB maps are made contain a wealth of temporal information. Modern ground-based experiments like the Atacama Cosmology Telescope (ACT) and the South Pole Telescope (SPT) have surveyed large patches of the sky for over a decade \cite{Chown:2018, ACT:Naess:2025}. The temporal information has been harnessed through the creation of ``time-division'' maps—a series of sky maps made from data binned in consecutive time intervals—which enable searches for time-varying phenomena.

This time-division approach has proven fruitful for time-domain astronomy, leading to the detection of transient sources like stellar flares and novae \cite{SPT:Whitehorn:2016, ACT:Naess:2021, SPT:Guns:2021, ACT:Li:2023, ACT:Biermann:2025} and enabling studies of Solar System objects \cite{ACT:Naess:2021:P9, SPT:Chichura:2022, ACT:Orlowski-Scherer:2024}. Beyond these astrophysical applications, time-division maps also provide a powerful probe of time-dependent cosmological signals. For instance, Ref.~\cite{Leluc:2025} recently proposed using these maps to detect a stochastic gravitational wave background (SGWB) passing Earth through the time-varying gravitational lensing effect it induces. Similarly, Ref.~\cite{Adachi:2023} employed time-division maps to search for time-dependent cosmic birefringence, a signature of ultralight axion-like particles that would cause a time-dependent rotation of the CMB's polarization.

Despite these promising advances, the time-division methodology is fundamentally constrained by its mapmaking cadence. To date, such analyses have been limited to using daily or multi-day maps, which restricts their frequency window to below $\sim 10^{-5}$\,Hz. This restriction leaves a vast frequency range—from microhertz to hertz—largely unexplored by CMB experiments for cosmological signals, leaving a discovery space open for signatures of new physics.

In this paper, we develop a new phase-folding technique to probe these oscillatory cosmological signals, drastically expanding the frequency reach of time-domain CMB science in this regime. Phase-folding is a standard technique in radio astronomy for studying pulsars, and has been extended to CMB data analysis in Ref.~\cite{Guan:2025} for detecting the Crab pulsar in ACT data. The key idea is to bin data according to the phase of a hypothetical periodic signal at a target frequency $f_s$, rather than by fixed time intervals. 
This shift from ``time-division'' to ``phase-division'' mapmaking overcomes the cadence limitation, enabling searches for periodic phenomena at up to the detector sampling rate of $\mathcal{O}(100)$\,Hz. In this work, we demonstrate that this technique can provide a powerful probe for cosmological signals across the entire sky, offering a clean way to isolate any periodic signal into frequency-specific modes that are orthogonal to the static primordial CMB.

We demonstrate the power of the phase-folding CMB map with two cosmological applications. First, we develop an estimator for detecting time-dependent lensing signal from SGWB passing Earth, extending the probe proposed in Ref.~\cite{Leluc:2025} above $10^{-5}$\,Hz. Second, we construct an estimator to search for an all-sky, oscillating polarization rotation signal---a signature of axion-like particles—which extends the CMB constraints on these particles into a higher mass range ($10^{-18} \lesssim m_a\lesssim 10^{-14}$\,eV).

This paper is organized as follows. In Sec.~\ref{sec:formalism}, we review the phase-folding mapmaking technique and the general formalism of its application to time-variable signals. In Sec.~\ref{sec:gw_app}, we detail its use for GW detection. In Sec.~\ref{sec:biref_app} we develop the estimator for time-dependent birefringence search. We conclude in Sec.~\ref{sec:conclusion}.

\section{The Phase-Folding Formalism}
\label{sec:formalism}
We begin by extending the phase-folding mapmaking technique introduced in Ref.~\cite{Guan:2025}. The core strategy is to transform the time-ordered data (TOD) from the time domain into a phase-frequency domain, where static astrophysical signals are cleanly separated from time-variable ones.

\subsection{Phase-Folding Mapmaking}
Consider a hypothetical, periodic signal of interest with a known frequency $f_s$ and corresponding period $P_s = 1/f_s$. The fundamental assumption of this method is that the total sky signal  can be decomposed into a series of static maps, $\mathbf{m}_b$, each corresponding to a specific phase of the periodic signal. The TOD from a detector $i$, denoted $d_i(t)$, can then be modeled as
\begin{equation}
    d_i(t) = \sum_{b=0}^{N_b-1} \eta_b(t) \left[ \mathbf{P}_{i}(t) \mathbf{m}_{b} + n_i(t) \right],
\end{equation}
where $\mathbf{P}_{i}(t)$ is the pointing matrix that maps sky pixels to the detector $i$ at time $t$, $\mathbf{m}_{b}$ is the sky map for phase bin $b$, and $n_i(t)$ is the detector noise. The phase-bin selection function, $\eta_b(t)$, is a periodic function that is unity when the phase falls within bin $b$ and zero otherwise,
\begin{equation}
    \eta_b(t) = 
    \begin{cases}
        1, & \text{if } \frac{b}{N_b} \leq \frac{(t-t_0) \bmod P_s}{P_s} < \frac{b+1}{N_b} \\
        0, & \text{otherwise.}
    \end{cases}
\end{equation}
where $t_0$ is a reference epoch for phase calculation. A crucial property of this function is the orthogonality of the bins: $\eta_b(t) \eta_{b'}(t) = \delta_{bb'} \eta_b(t)$. This orthogonality ensures that the data corresponding to each phase bin are independent, allowing one to solve for each $\mathbf{m}_b$ separately. The maximum-likelihood solution is given by
\begin{equation}
  \hat{\mathbf{m}}_b = \left( \mathbf{P}^T \boldsymbol{\eta}_b \mathbf{N}^{-1} \boldsymbol{\eta}_b \mathbf{P} \right)^{-1} \mathbf{P}^T \boldsymbol{\eta}_b \mathbf{N}^{-1} \mathbf{d},
  \label{eq:m_b}
\end{equation}
where $\mathbf{d}$ is the full TOD, $\mathbf{N}$ is the noise covariance matrix, and $\mathbf{\eta}_b$ is a diagonal matrix with elements $[\boldsymbol{\eta}_b]_{tt} = \eta_b(t)$. The output of this process is a set of $N_b$ independent sky maps, $\{\mathbf{m}_b\}$, each representing the average sky signal observed during a specific phase of the periodic signal.

An important parameter in this framework is the number of phase bins, $N_b$. This choice is constrained by the detector sampling rate, $f_{\text{sample}}$, and the target signal frequency, $f_s$, as the number of bins must be less than the number of samples per oscillation period, i.e., $N_b \leq f_{\text{sample}}/f_s$. This introduces a practical trade-off. To probe the theoretical maximum frequency, the Nyquist limit of $f_{\text{sample}}/2$, one must use the minimum number of bins, $N_b = 2$. However, as we will show in Sec.~\ref{sec:gw_app}, averaging the signal over a finite bin width introduces a window function, $W(N_b) = \text{sinc}(\pi/N_b)$, which suppresses the signal amplitude. For $N_b = 2$, this suppression is significant ($W(2) \approx 0.64$), while a larger number of bins like $N_b = 10$ provides high signal fidelity ($W(10) \approx 0.98$) at the cost of a lower maximum frequency ($f_s \leq f_{\text{sample}}/N_b$). While our framework extends sensitivity to the Nyquist limit, in practice one must balance this trade-off to maximize signal-to-noise on the target frequency range.

\subsection{Frequency Modes}
\label{subsec:dft}
With the set of phase-binned maps, $\{m_b(\hat{\mathbf{n}})\}$, in hand, we can isolate a signal oscillating at frequency $f_s$ by applying a Discrete Fourier Transform (DFT) across the phase-bin dimension. We define the frequency-mode maps $\mathcal{M}(\hat{\mathbf{n}}, k)$ as
\begin{equation}
    \mathcal{M}(\hat{\mathbf{n}}, k) = \frac{1}{N_b} \sum_{b=0}^{N_b-1} \mathbf{m}_b(\hat{\mathbf{n}}) e^{-i 2\pi b k / N_b},
    \label{eq:dft_maps}
\end{equation}
where $k \in \{0, 1, \dots, N_b-1\}$ is the discrete frequency index. It is more convenient to work in harmonic space. We define the spherical harmonic coefficients of the $k$-th frequency-mode map, $\mathcal{T}_{\ell m}(k)$, as
\begin{equation}
\begin{split}
    \mathcal{T}_{\ell m}(k) &= \int d^2\mathbf{\hat{n}} \, \mathcal{M}(\hat{\mathbf{n}}, k) Y_{\ell m}^*(\hat{\mathbf{n}}) \\
    &= \frac{1}{N_b} \sum_{b=0}^{N_b-1} T_{\ell m, b} \, e^{-i 2\pi b k / N_b},
\end{split}
\end{equation}
where $T_{\ell m, b}$ are the spherical harmonic coefficients of the map for phase bin $b$. These frequency-mode coefficients have a clear physical interpretation. The $k=0$ mode, $\mathcal{T}_{\ell m}(0)$, represents the average of all phase-binned maps and thus contains the static sky, including the primordial CMB, galactic foregrounds, and any other non-time-varying signals. The $k=1$ mode, $\mathcal{T}_{\ell m}(1)$, is the fundamental mode and isolates any signal component that oscillates sinusoidally at the folding frequency $f_s$. Higher modes with $k>1$ capture harmonics of the fundamental frequency ($k \cdot f_s$), which would be present for periodic signals with non-sinusoidal profiles. Because the input maps are real, these coefficients obey the conjugate symmetry $\mathcal{T}_{\ell m}(k) = \mathcal{T}_{\ell m}(N_b-k)^*=\mathcal{T}_{\ell m}(-k)^*$.

\subsection{Statistical Properties of the Frequency Modes}
\label{subsec:stats}
The power of this formalism comes from the distinct statistical properties of the different frequency modes. Let us model the true sky signal in each phase-binned map in harmonic space as $T_{\ell m, b} = S_{\ell m} + \delta T_{\ell m, b} + n_{\ell m, b}$, where $S_{\ell m}$ is the static sky, $\delta T_{\ell m, b}$ is a small time-variable signal component, and $n_{\ell m, b}$ is the instrumental noise in that bin's map.

Since each of the $N_b$ phase bins uses only $1/N_b$ of the total integration time, the instrumental noise power in a single binned map is $N_b$ times larger than the noise in a map made from the full dataset. Assuming the noise is uncorrelated between bins, the noise covariance in harmonic space is
\begin{equation}
    \langle n_{\ell_1 m_1, b_1} n_{\ell_2 m_2, b_2}^* \rangle = \delta_{b_1 b_2} \delta_{\ell_1 \ell_2} \delta_{m_1 m_2} (N_b N_{\ell_1}^{\text{total}}),
\end{equation}
where $N_\ell^{\text{total}}$ is the noise power spectrum of a map constructed from the entire survey data, assuming statistical isotropy.

We now compute the covariance of the frequency-mode coefficients, $\mathcal{T}_{\ell m}(k)$, in the absence of a time-variable signal ($\delta T_{\ell m, b} = 0$). For the static mode ($k=0$),
\begin{equation}
    \mathcal{T}_{\ell m}(0) = \frac{1}{N_b} \sum_{b=0}^{N_b-1} \left( S_{\ell m} + n_{\ell m, b} \right) = S_{\ell m} + \frac{1}{N_b}\sum_{b=0}^{N_b-1} n_{\ell m, b}.
\end{equation}
The power spectrum of this mode is the sum of the static signal power and the averaged noise power
\begin{equation}
  \langle \mathcal{T}_{\ell_1 m_1}(0) \mathcal{T}_{\ell_2 m_2}^*(0) \rangle = \delta_{\ell_1 \ell_2} \delta_{m_1 m_2} \left( \tilde{C}_{\ell_1} + N_{\ell_1}^{\text{total}} \right),
\end{equation}
where $\tilde{C}_\ell = C_\ell^{\text{CMB}} + C_\ell^{\text{fg}}$ includes both CMB and static foreground power. This is the familiar power spectrum of a standard, full-mission coadded map.

For the oscillating modes ($k \neq 0$),
\begin{equation}
    \mathcal{T}_{\ell m}(k) = \frac{1}{N_b} \sum_{b=0}^{N_b-1} \left( S_{\ell m} + n_{\ell m, b} \right) e^{-i 2\pi b k / N_b}.
\end{equation}
The static signal term, $S_{\ell m}$, vanishes because $\sum_b e^{-i 2\pi b k / N_b} = 0$ for any non-zero integer $k < N_b$. This leaves only the contribution from instrumental noise
\begin{align}
    &\langle \mathcal{T}_{\ell_1 m_1}(k) \mathcal{T}_{\ell_2 m_2}^*(k') \rangle \\
    &\quad = \frac{1}{N_b^2} \sum_{b,b'} e^{-i 2\pi (bk - b'k')/N_b} \langle n_{\ell_1 m_1, b} n_{\ell_2 m_2, b'}^* \rangle \nonumber \\
    &\quad = \frac{\delta_{\ell_1 \ell_2} \delta_{m_1 m_2}}{N_b^2} \sum_{b} e^{-i 2\pi b (k-k')/N_b} (N_b N_{\ell_1}^{\text{total}}) \nonumber \\
    &\quad = \delta_{\ell_1 \ell_2} \delta_{m_1 m_2} \delta_{kk'} N_{\ell_1}^{\text{total}}.
    \label{eq:noise_k_neq_0}
\end{align}
This result shows that for any non-zero frequency mode ($k \neq 0$), the power spectrum contains only instrumental noise. The primordial CMB, static foregrounds, and their associated cosmic variance are entirely confined to the $k=0$ mode. This provides a cosmic-variance-free space to search for faint, time-variable signals, which will appear as power or correlations in the $k \neq 0$ modes. We also note that the noise power spectrum, $N_{\ell}^{\text{total}}$, is identical to the static $k=0$ mode. This implies that the method has uniform sensitivity to a signal at any frequency $f_s$. This is not surprising, as the phase binning divides the data into $N_b$ subsets, and the DFT redistributes the total noise from these bins evenly across the frequency modes.

\section{Time-dependent deflection from Gravitational Waves}
\label{sec:gw_app}
\begin{figure*}[t]
  \centering
\includegraphics[width=0.8\linewidth]{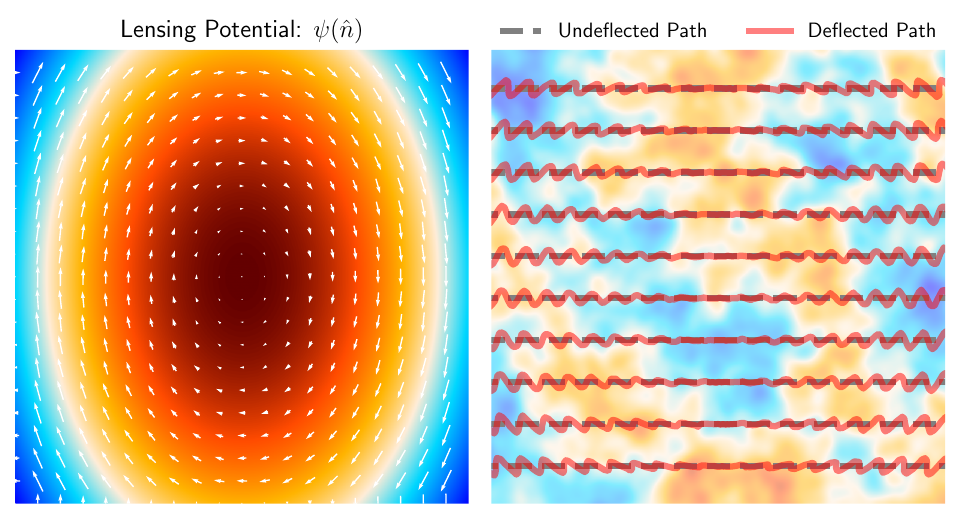}
    \caption{The left panel shows the lensing potential $\psi$ used in the toy model on a $10^\circ \times 10^\circ$ patch. The white arrows illustrate the direction and relative size of $\star\nabla \psi$ (scaled for visualization). The right panel illustrates the effect of a 10\,Hz gravitational background to the effective scan path. The black dashed curve illustrates the undeflected scan pattern; the red curves illustrates the deflected scan pattern due to lensing effect from the 10\,Hz gravitational wave. While all pixels are scanned across in an observation, we only show selected scan paths for illustration. The effect of deflection has been enlarged by 50 times in this visualization. The background shows the primary CMB field.}
    \label{fig:deflection_illustration}
\end{figure*}

The search for gravitational waves (GWs) provides a unique window onto the universe, with different techniques probing distinct parts of the GW frequency spectrum. Ground-based interferometers such as LIGO \cite{LIGOScientificCollaboration:2015} and Virgo \cite{Acernese:2015} have opened the high-frequency band from 10 to $10^4$\,Hz. The future space-based interferometer LISA \cite{Colpi:2024} is designed to target the millihertz range ($10^{-4}$ to 1\,Hz), while Pulsar Timing Arrays (PTAs) are currently placing constraints in the nanohertz regime ($10^{-9}$ to $10^{-7}$\,Hz) \cite{Agazie:2023}. In addition to these methods, searches for primordial GWs from inflation seek a characteristic B-mode polarization pattern in the CMB \cite{KKS:1997, Zaldarriaga:1997}.

An alternative approach to detecting GWs is through high-precision astrometry. A GW passing Earth induces a time-varying deflection in the apparent positions of all distant sources \cite{Book:2011}. This astrometric signature is being actively pursued using data from observatories like Gaia \cite{GaiaCollaboration:2016}, which has been used to constrain stochastic GW backgrounds in a frequency range complementary to PTAs \cite{Darling:2018, Jaraba:2023}.

This same phenomenon of time-dependent gravitational lensing also affects CMB photons. Ref.~\cite{Leluc:2025} recently demonstrated that this effect can be detected using time-resolved CMB maps. However, their approach, which relied on daily mapmaking, was limited to frequencies below the $\sim 10^{-5}$\,Hz cadence. In this section, we apply the phase-folding formalism from Sec.~\ref{sec:formalism} to extend the search for GWs to much higher frequencies. This method transforms CMB experiments into broadband GW probes, opening a new observational window from millihertz to hertz scales.

\subsection{Time-Dependent Lensing from a GW}
\label{subsec:gw_signal}
A passing GW perturbs the spacetime metric, leading to a time-dependent deflection of the apparent positions of CMB photons. This deflection field, $\mathbf{d}(\hat{\mathbf{n}}, t)$, can be decomposed into gradient ($\phi$) and curl ($\psi$) potentials,
\begin{equation}
\mathbf{d}(\hat{\mathbf{n}}, t) = \nabla \phi(\hat{\mathbf{n}}, t) + \star \nabla \psi(\hat{\mathbf{n}}, t),
\end{equation}
where $\star$ denotes a rotation of the gradient vector by 90$^\circ$. While both components are generated, the curl mode $\psi$ is of particular interest for GW detection, as it receives negligible contamination from the dominant, static lensing by large-scale structure.

We consider a GW with frequency $f_{\text{gw}}$. The time-dependent deflection potential at each point on the sky, $\Gamma(\hat{\mathbf{n}}, t)$, can be expanded in spherical harmonics as $\Gamma(\hat{\mathbf{n}}, t) = \sum_{\ell m} \Gamma_{\ell m}(t) Y_{\ell m}(\hat{\mathbf{n}})$. We model the time evolution of these harmonic coefficients can be written as
\begin{equation}
\Gamma_{\ell m}(t) = [\Gamma_{0}]_{\ell m} \cos(\omega_{\text{gw}} t + \delta),
\label{eq:gw_potential_time_lm}
\end{equation}
where $\Gamma$ represents either $\phi$ or $\psi$, $\omega_{\text{gw}} = 2\pi f_{\text{gw}}$, $[\Gamma_{0}]_{\ell m}$ are the spherical harmonic coefficients of $\Gamma_0(\hat{\mathbf{n}})$, and $\delta$ is the phase of the GW.

To isolate this signal using our formalism, we fold the data at the GW frequency, $f_s = f_{\text{gw}}$. The spherical harmonic coefficients of the deflection potential averaged within phase bin $b$ (with central phase $\varphi_b = 2\pi b / N_b$) are then
\begin{equation}
\Gamma_{\ell m, b} = [\Gamma_{0}]_{\ell m} \cos(\varphi_b + \delta) W(N_b),
\end{equation}
where $W(N_b) = \text{sinc}(\pi/N_b)$ is a window function that accounts for averaging a sinusoid over a finite phase bin. For a sufficient number of bins ($N_b \gg 1$), this factor approaches unity.

Applying the DFT from Eq.~\eqref{eq:dft_maps}, the component at $k=1$ is directly related to the amplitude of the GW signal by
\begin{equation}
\begin{split}
\Gamma_{\ell m}(k=1) &= \frac{1}{N_b} \sum_{b=0}^{N_b-1} \Gamma_{\ell m, b} e^{-i 2\pi b / N_b}, \\
&= [\Gamma_0]_{\ell m} e^{i\delta} \frac{W(N_b)}{2}.
\end{split}
\label{eq:xi_k1}
\end{equation}
The $k=-1$ (or $k=N_b-1$) mode is simply its complex conjugate and contains no additional information. This equation provides a direct link between an observable quantity, $\Gamma_{\ell m}(k=1)$, and the amplitude of the gravitational wave, $\Gamma_0$.

\subsection{Quadratic Estimator}
\label{subsec:gw_estimator}
For notational simplicity, we now transition from full-sky formalism to flat-sky approximation. This approximation is valid for analyzing small patches of the sky or in the high-multipole limit ($\ell \gg 1$).\footnote{In the flat-sky limit, spherical harmonic transforms $\sum_{\ell m}$ become 2D Fourier integrals $\int d^2\mathbf{l}/(2\pi)^2$, spherical harmonics $Y_{\ell m}$ become plane waves $e^{i\mathbf{l}\cdot\mathbf{x}}$.}

Gravitational lensing remaps the primordial CMB temperature field, $\tilde{T}$, to $T^{\text{obs}}=\tilde{T}(\mathbf{n} + \mathbf{d})$, with $\mathbf{d}$ being the deflection field. To first order in the deflection and in the flat-sky limit, the observed temperature in phase bin $b$, $T^{\text{obs}}_b$, is given by
\begin{equation}
\begin{aligned}
T^{\mathrm{obs}}_b(\mathbf{l})
&\approx \tilde{T}(\mathbf{l})
- \int \frac{d^2\mathbf{l}'}{(2\pi)^2}\,\tilde{T}(\mathbf{l}')\\
&\qquad\times\Big[ (\mathbf{l}'\!\cdot\!\mathbf{L})\,\phi_b(\mathbf{L})
+ (\mathbf{l}'\!\cdot\!\star\mathbf{L})\,\psi_b(\mathbf{L}) \Big],
\end{aligned}
\end{equation}
where $\mathbf{L} = \mathbf{l} - \mathbf{l}'$, and $\phi_b(\mathbf{L})$ and $\psi_b(\mathbf{L})$ are the averaged deflection potential $\phi(\mathbf{L}, t)$, $\psi(\mathbf{L},t)$ in phase bin $b$, respectively.

\begin{figure*}[t]
\centering
\includegraphics[width=\linewidth]{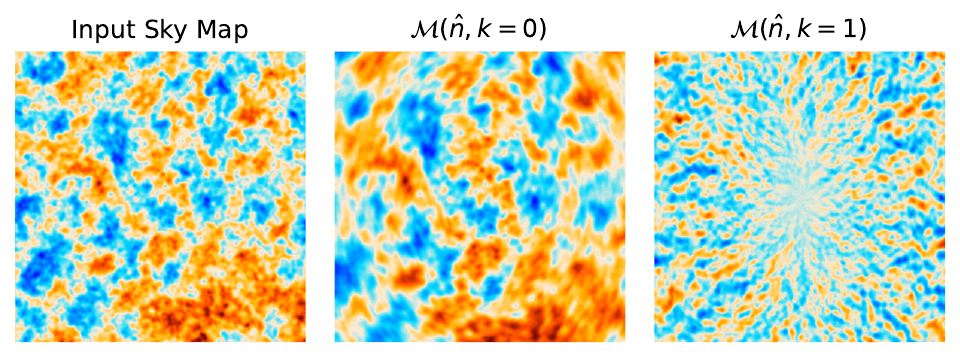}
\caption{Illustration of the phase-folding mapmaking output for the toy model. \textbf{Left:} The input, unlensed CMB temperature anisotropy map on the $8^\circ \times 8^\circ$ patch. \textbf{Middle:} The average sky map, $\mathcal{M}(\hat{\mathbf{n}}, 0)$, constructed from all data, showing the standard CMB. The effect of the time-dependent lensing causes a blurry effect when exaggerated by 50 times for visualization purpose. \textbf{Right:} The oscillating mode map, $\mathcal{M}(\hat{\mathbf{n}}, 1)$, which isolates the signal modulated at the GW frequency. The clear, non-random spatial pattern is indicative of a coherent signal.}
\label{fig:input_output}
\end{figure*}

Transforming to the phase-frequency domain using Eq.~\eqref{eq:dft_maps}, and noting that the unlensed CMB is static ($\tilde{T}(\mathbf{l}, k) = \tilde{T}(\mathbf{l}) \delta_{k0}$), we find the frequency modes of the observed temperature
\begin{equation}
\begin{aligned}
\mathcal{T}^{\text{obs}}(\mathbf{l}, k) &\approx \tilde{T}(\mathbf{l}) \delta_{k0} - \int \frac{d^2\mathbf{l}'}{(2\pi)^2} \tilde{T}(\mathbf{l}') \\
&\quad\times \left[ (\mathbf{l}' \cdot \mathbf{L})\phi(\mathbf{L}, k) + (\mathbf{l}' \cdot \star\mathbf{L})\psi(\mathbf{L}, k) \right],
\end{aligned}
\label{eq:lensed_T_general}
\end{equation}
where
\begin{equation}
\begin{aligned}
\phi(\mathbf{L}, k) &\equiv \frac{1}{N_b}\sum_b \phi_b(\mathbf{L}) e^{-i2\pi bk/N_b},\\
\psi(\mathbf{L}, k) &\equiv \frac{1}{N_b}\sum_b \psi_b(\mathbf{L}) e^{-i2\pi bk/N_b}.
\end{aligned}
\end{equation}
Eq.~\eqref{eq:lensed_T_general} suggests that a time-dependent deflection mode at $k=1$ scatters power from the static, unlensed CMB at $k=0$ to the lensed CMB at $k=1$. This induces a specific off-diagonal covariance between the observed temperature modes,
\begin{equation}
\begin{aligned}
\langle \mathcal{T}^{\text{obs}}(\mathbf{l}_1, 0) \mathcal{T}^{\text{obs}*}(\mathbf{l}_2, 1) \rangle &= f^\phi(\mathbf{l}_1, \mathbf{l}_2) \phi(\mathbf{L}, 1) \\
&\quad + f^\psi(\mathbf{l}_1, \mathbf{l}_2) \psi(\mathbf{L}, 1),
\end{aligned}
\label{eq:off_diag_corr}
\end{equation}
where $\mathbf{L} = \mathbf{l}_2 - \mathbf{l}_1$, and the mode-coupling kernels for the gradient and curl components are
\begin{align}
f^\phi(\mathbf{l}_1, \mathbf{l}_2) &= -[\mathbf{l}_1 \cdot \mathbf{L}] C_{l_1}^{TT}, \\
f^\psi(\mathbf{l}_1, \mathbf{l}_2) &= -[\mathbf{l}_1 \cdot \star \mathbf{L}] C_{l_1}^{TT}.
\end{align}

We can construct an unbiased quadratic estimator for a general deflection potential $\Gamma \in \{\phi, \psi\}$. Following the notation in \cite{Leluc:2025}, we introduce a generalized product operator $\odot$ such that $\mathbf{a} \odot \mathbf{b} = \mathbf{a} \cdot \mathbf{b}$ for the gradient mode ($\Gamma=\phi$) and $\mathbf{a} \odot \mathbf{b} = \mathbf{a} \cdot \star \mathbf{b}$ for the curl mode ($\Gamma=\psi$). The general estimator for the $k=1$ deflection mode is then
\begin{equation}
\hat{\Gamma}(\mathbf{L}, 1) = A^\Gamma(L) \int \frac{d^2\mathbf{l}_{1}}{(2\pi)^2} \mathcal{T}^{\text{obs}}(\mathbf{l}_{1}, 0) \mathcal{T}^{\text{obs}*}(\mathbf{l}_2, 1) F^\Gamma(\mathbf{l}_1, \mathbf{l}_2),
\label{eq:gen_estimator}
\end{equation}
where $\mathbf{l}_2 = \mathbf{L} + \mathbf{l}_1$. The optimal filter $F^\Gamma$, which minimizes the variance, is given by
\begin{equation}
F^\Gamma(\mathbf{l}_1, \mathbf{l}_2) = \frac{f^\Gamma(\mathbf{l}_1, \mathbf{l}_2)^*}{C_{l_1}^{\text{obs}}(0) C_{l_2}^{\text{obs}}(1)} = \frac{-[\mathbf{l}_{1} \odot \mathbf{L}] C_{l_1}^{TT}}{\left[C_{l_1}^{TT} + N_{l_1}^{\text{total}}\right] N_{l_2}^{\text{total}}}.
\label{eq:gen_filter}
\end{equation}
Here, $C_{l}^{\text{obs}}(0) = C_{l}^{TT} + N_{l}^{\text{total}}$ is the total power in the static ($k=0$) mode,\footnote{We have neglected the contribution from foregrounds here as it is not essential to our methodology here. Including the foreground power spectrum in $C_l^{\rm obs}(0)$ would further enhance the sensitivity of this quadratic estimator, as shown in \cite{Leluc:2025}.} and $C_{l}^{\text{obs}}(1) = N_{l}^{\text{total}}$ is the noise power in the oscillating ($k=1$) mode. The normalization $A^\Gamma(L)$ ensures the estimator is unbiased (i.e., $\langle \hat{\Gamma}(\mathbf{L}, 1) \rangle = \Gamma(\mathbf{L}, 1)$) and is given by
\begin{equation}
A^\Gamma(L) = \left[ \int \frac{d^2\mathbf{l}_1}{(2\pi)^2} f^\Gamma(\mathbf{l}_1, \mathbf{l}_2) F^\Gamma(\mathbf{l}_1, \mathbf{l}_2)^* \right]^{-1}.
\end{equation}
The variance of this estimator in the absence of a signal defines the reconstruction noise power spectrum
\begin{equation}
\mathcal{N}^{\Gamma\Gamma}_{L} \equiv \langle |\hat{\Gamma}(\mathbf{L}, 1)|^2 \rangle_{\text{no signal}} = A^\Gamma(L).
\label{eq:recon_noise_gen}
\end{equation}
One can further show that the noise spectra for the gradient and curl modes are identical, i.e., $\mathcal{N}^{\phi\phi}_{L} = \mathcal{N}^{\psi\psi}_{L}$, and their forms are identical to the result derived in \cite{Leluc:2025} with the time-division framework. This confirms that the phase-folding method effectively redistributes the statistical power of the full survey without altering the fundamental noise floor.

\subsection{A Toy Model}
\label{subsec:gw_toy}

\begin{figure*}[t]
\centering
\includegraphics[width=\linewidth]{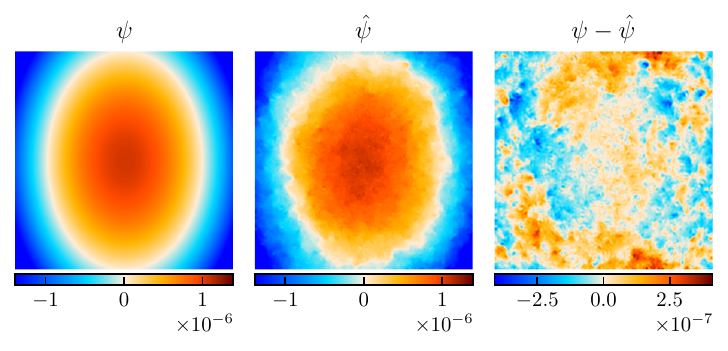}
\caption{Demonstration of the phase-resolved lensing reconstruction in a toy model. \textbf{Left:} The input curl potential $\psi(\hat{\mathbf{n}})$ (a pure $l=2, m=2$ mode) used to generate the time-dependent deflection. \textbf{Middle:} The reconstructed curl potential $\hat{\psi}(\hat{\mathbf{n}})$ recovered by applying the quadratic estimator (Eq.~\eqref{eq:gen_estimator}) to the simulated data. \textbf{Right:} The residual difference map, $\psi - \hat{\psi}$, which is consistent with instrumental noise. The successful recovery of the input signal validates the formalism.}
\label{fig:psi_diff}
\end{figure*}

We build a toy model to demonstrate the methodology with a proof-of-concept simulation on a flat sky. While the quadratic estimator formalism is general, we specialize to the reconstruction of the curl potential $\psi$ for this demonstration. 
In the context of our time-dependent formalism, which correlates a static ($k=0$) map with an oscillating ($k=1$) map, both the $\phi$ and $\psi$ estimators are effective probes. The method inherently isolates time-varying signals and is not biased by the static lensing potential from large-scale structure. We therefore choose $\psi$ as a clear, representative example for this simulation, but the formalism applies equally to $\phi$. Our goal is to demonstrate the recovery of a known, time-varying lensing potential from simulated time-ordered data using the phase-folding and quadratic estimator approach.

We begin by generating a Gaussian realization of the primary (unlensed) CMB temperature anisotropy on a high-resolution $10^\circ \times 10^\circ$ patch with $0.1'$ pixels, based on the best-fit \textit{Planck} 2018 power spectrum \cite{Planck2018:VI:CP}. This map is convolved with a symmetric Gaussian beam with $\theta_{\rm fwhm} = 1.4'$, representative of ground-based experiments like ACT or the SO Large Aperture Telescope (LAT). The time-variable signal is sourced by a curl-only GW potential. For clear visualization, we choose a specific spatial pattern, $\psi(\hat{\mathbf{n}})$, corresponding to a pure $Y_{22}$ spherical harmonic, as visualized in the left panel in Fig.~\ref{fig:deflection_illustration}. The amplitude is set such that the root-mean-square deflection is approximately $10$ arcseconds, which is artificially chosen for demonstration purposes. This potential oscillates monochromatically at $f_{\text{gw}} = 10$\,Hz, producing a deflection field $\mathbf{d}(\hat{\mathbf{n}}, t) = \star\nabla\psi(\hat{\mathbf{n}})\cos(2\pi f_{\text{gw}} t)$.

We simulate a total of 100 observations each with a single detector scanning an $8^\circ \times 8^\circ$ patch with $0.5'$ pixels. To ensure uniform coverage, the scan pattern alternates between left-to-right and up-to-down boustrophedon (raster) paths. For each of 100 simulated scans, we generate a time-ordered data stream sampled at 200\,Hz. The signal in the TOD is calculated by interpolating the beam-convolved, high-resolution CMB map at the lensed position, $\hat{\mathbf{n}}_{\text{sky}} = \hat{\mathbf{n}}_{\text{point}}(t) - \mathbf{d}(\hat{\mathbf{n}}, t)$, using bilinear interpolation, as illustrated in the right panel of Fig.~\ref{fig:deflection_illustration}. We then add Gaussian white noise of $4\,\mu\text{K}\sqrt{\text{s}}$ to the timestream. This noise level is chosen such that the coadded map noise after all 100 scans is approximately $4\,\mu\text{K}'$. We neglect atmospheric $1/f$ noise in this toy example, which we leave for future work.

The simulated TODs are folded at the 10\,Hz signal frequency and binned into $N_b=10$ separate phase maps, $\{m_b(\hat{\mathbf{n}})\}$ using Eq.~\eqref{eq:m_b}. The frequency-mode maps, $\mathcal{M}(\hat{\mathbf{n}}, k)$, are then computed via a DFT (Eq.~\eqref{eq:dft_maps}). The resulting maps are shown in Fig.~\ref{fig:input_output}. The static mode ($k=0$) contains the standard CMB sky, while the oscillating mode ($k=1$) clearly reveals a coherent spatial pattern distinct from noise.

Finally, we apply the quadratic estimator from Eq.~\eqref{eq:gen_estimator} to these frequency-mode maps, using the \texttt{symlens} library. We correlate the static map $\mathcal{M}(\hat{\mathbf{n}}, 0)$ with the oscillating map $\mathcal{M}(\hat{\mathbf{n}}, 1)$, using multipoles in the range $30 \leq l \leq 3000$ and applying a $0.1^\circ$ apodization to the analysis mask. The resulting estimate, $\hat{\psi}(\mathbf{L}, 1)$, recovers the complex quantity $\psi_0(\mathbf{L})e^{i\delta}\frac{W(N_b)}{2}$. To obtain the true amplitude, we correct for the phase-binning window function: $\hat{\psi}_0 = \frac{2}{W(N_b)}\hat{\psi}(\mathbf{L}, 1)$, where $W(N_b) \approx 0.984$ for $N_b=10$. Fig.~\ref{fig:psi_diff} shows the results. The left panel displays the input curl potential, the middle panel shows our reconstructed potential, and the right panel shows the residual difference. The successful recovery of the input signal and the noise-like character of the residual validate the end-to-end formalism.

\subsection{Sensitivity}
\label{subsec:gw_sensitivity}

\begin{figure}[t]
\centering
\includegraphics[width=\linewidth]{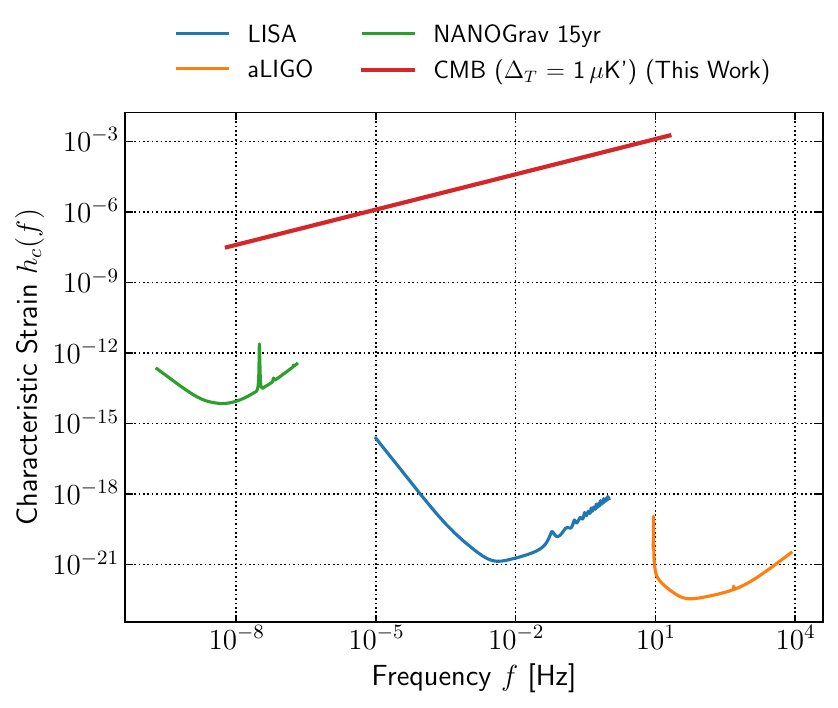}
\caption{Forecasted sensitivity to the characteristic strain $h_c(f)$ of a stochastic gravitational wave background. The red curve shows the projected reach of the phase-folding CMB method with SO LAT-like experiment with $\Delta_T=1\,\mu \mathrm{K}'$, $f_{\text{sky}}=0.5$, $N_b=10$, integrated over 10 years with a 200\,Hz sampling rate. We compare against sensitivity curves from pulsar timing arrays (NANOGrav 15 year, green), space-based interferometers (LISA, blue, and ground-based detectors (Advanced LIGO, orange).}
\label{fig:sensitivity_curve}
\end{figure}

With the estimator formalism demonstrated, we can forecast the
sensitivity of this method to SGWB. A statistically isotropic SGWB is often characterized by its characteristic strain, $h_{c}(f)$, which is related to its energy density spectrum $\Omega_{\rm GW}(f)$ as
\begin{equation}
h_{c}(f)^{2} = \frac{3H_{0}^{2}\Omega_{\rm GW}(f)}{2\pi^{2}f^{2}}
\end{equation}
where $H_0$ is the Hubble constant. $h_{c}(f_{\rm gw})$ sources a time-dependent deflection field at a GW frequency $f_{\text{gw}}$, with power spectral density given by \cite{Book:2011, Leluc:2025},
\begin{equation}
C_L^{\Gamma\Gamma,\text{GW}}(f_{\text{gw}}) = g_\Gamma \alpha_L^{\Gamma\Gamma} \frac{h_c^2(f_{\text{gw}})}{6 f_{\text{gw}} L^3},
\label{eq:signal_power_hc}
\end{equation}
where $\Gamma \in \{\phi, \psi\}$. The geometrical factors
$\alpha_L^{\Gamma\Gamma}$ encode the angular dependence of the
deflection, which is predominantly quadrupolar ($L=2$). A key feature
of GW-induced deflection is that it sources gradient and curl modes
with equal power, i.e., $g_\phi = g_\psi = 1/2$. Note that the
quadratic estimator as defined in Eq.~\eqref{eq:gen_estimator}
directly measures
$\Gamma(\mathbf{L}, 1) = \frac{W(N_b)}{2}\Gamma_0(\mathbf{L})e^{i\delta}$,
which includes the window function suppression. For typical choices of
$N_b \gtrsim 10$, we have $W(N_b) > 0.98$, making this correction
small. In what follows, we absorb this factor into the definition of
the signal amplitude for simplicity, noting that a correction of
$[W(N_b)]^{-2} \approx 1.04$ should be applied to the power spectrum
when $N_b = 10$.

Our method provides an estimator, $\hat{C}_L^{\Gamma\Gamma,\text{GW}}(f)$, targeting a given frequency, $f_{\text{gw}}$, integrated over a finite frequency bin width of $1/\Delta t_{\rm survey}$, with $\Delta t_{\rm survey}$ being the total duration of the survey. The variance of the estimator is given by
\begin{equation}
  \text{Var}\left[\hat{C}_{L}^{\rm \Gamma\Gamma,GW}(f_{\rm gw})\right] = \frac{2}{(2L+1)f_{\rm sky}}\left(\Delta t_{\rm survey}\mathcal{N}_{\rm L}^{\Gamma\Gamma}\right)^{2},
\end{equation}
where $\Delta t_{\rm survey} \mathcal{N}_{\rm L}^{\Gamma\Gamma}$ represents
the effective instantaneous reconstruction noise power, and $f_{\rm sky}$ is the fraction of sky coverage. This result is consistent with \cite{Leluc:2025}. Therefore we expect the same sensitivity as the time-resolved approach while dramatically extending the frequency coverage up to $\mathcal{O}(10\,\text{Hz})$.

As the deflection power is predominantly in the largest angular scales
(quadrupolar; $L=2$), we can take the limit that $|\mathbf{l}| \gg L$, in which
case the reconstruction noise becomes approximately scale-independent
\begin{equation}
  \begin{split}
    L^2 \mathcal{N}^{\Gamma\Gamma}_{L} &\approx \left[ \frac{1}{2\pi} \int_0^\infty dl\; \frac{l^{3}(\tilde{C}_{l}^{TT})^2}{C_{l}^{\rm TT,obs} N_{l}^{\text{total}}} \right]^{-1}.
  \end{split}
  \label{eq:noise_approx}
\end{equation}
Here $\tilde{C}_{l}^{TT}$ refers to the undeflected power spectrum,
and $C_{l}^{\rm TT,obs} = C_{l}^{\rm TT} + N_{l}^{\rm total}$. Note
that the reconstruction noise is independent of the target GW
frequency, $f_{\rm gw}$, which is not surprising since the
phase-folding maps contain equal amount of statistics for any target
frequency. We model the instrument noise as
\begin{equation}
  N_{l} = \Delta_{T}^{2}\exp\left(l(l+1)\frac{\theta_{\rm fwhm}^{2}}{8\ln 2}\right),
\end{equation}
where $\Delta_{T}$ is the white noise level of the integrated map,
$\theta_{\rm fwhm}$ is the full-width-half-maximum of the instrument
beam. For a large-aperture CMB experiment with
$\theta_{\rm fwhm}=1.4'$,
\begin{equation}
  L^{2}\mathcal{N}_{L}^{\Gamma\Gamma} \approx 4.08\times 10^{-17} \left[\frac{\Delta_{T}}{1\mu \mathrm{K}'}\right]^{2}
\end{equation}
in the absence of astrophysical foreground. While including
astrophysical foreground is expected to further enhance our
sensitivity \cite{Leluc:2025}, it is not the main focus of our
discussion.

It is useful to express GW power in terms power spectral density, $S_{h}(f) \equiv h_{c}(f)^{2}/f$. Our estimator gives,
\begin{equation}
  \hat{S}_{h}(f) = \frac{6L^{3}}{g_{\Gamma}\alpha_{L}^{\Gamma\Gamma}}\hat{C}_{L}^{\rm \Gamma\Gamma,GW}(f_{\rm gw}),
\end{equation}
using a specific $L$ and $\Gamma$ mode. It has a frequency independent variance given by
\begin{equation}
  \begin{split}
    \text{Var}\!\left[\hat{S}_{h}(f)\right]_{\Gamma,L}
    &= \left[\frac{72L^{6}}{(2L+1)f_{\rm sky}(g_{\Gamma}\alpha_{L}^{\Gamma\Gamma})^{2}} \right]
      \\
    &\quad\times \left(\Delta t_{\rm survey} \, \mathcal{N}_{L}^{\Gamma\Gamma}\right)^{2}.
  \end{split}
\end{equation}
We can make a minimum variance estimator, $\hat{S}_{h, \text{mv}}(f)$ by combining the statistics from $\Gamma \in \psi, \phi$ and all $L$ modes with inverse variance weighting. The variance of this minimum variance estimator is given by
\begin{equation}
  \begin{split}
    \text{Var}\left[\hat{S}_{h,\text{mv}}(f_{\rm gw})\right]
    &= \left[\sum_{\Gamma \in \phi, \psi}\sum_{L=2}^{\infty}\frac{(2L + 1) f_{\rm sky}(g_{\Gamma}\alpha_{L}^{\Gamma\Gamma})^{2}}{72 L^{6} \left(\Delta t_{\rm survey} \, \mathcal{N}_{L}^{\Gamma\Gamma}\right)^{2}}\right]^{-1}.
  \end{split}
\end{equation}
which is a frequency independent constant for each experiment setup. Consider only the quadrupolar mode ($L=2$) with $\alpha^{\Gamma\Gamma}_{2}=5/6$, $g_{\phi}=g_{\psi}=1/2$, and $N_{L}^{\phi\phi}=N_{L}^{\psi\psi}$, we get
\begin{equation}
  \text{Var}\left[S_{h,\text{mv}}(f_{\rm gw})\right] \approx 5.0\times 10^{-14}\left[\frac{\Delta_{T}}{1\mu \mathrm{K}'}\right]^{4}\left[\frac{0.5}{f_{\rm sky}}\right]\left[\frac{\Delta t_{\rm survey}}{10\,\text{yr}}\right]^{2}
\end{equation}
for a SO LAT-like experiment with $\theta_{\rm fwhm}=1.4'$ using CMB alone (no astrophysical foreground). The sensitivity to characteristic strain, $h_c(f)$, can then be approximated as
\begin{equation}
\sigma[h_c(f)] = \left(f^2\text{Var}[S_{h,\text{mv}}]/2\right)^{1/4} \propto f^{1/2}.
\end{equation}

In Fig.~\ref{fig:sensitivity_curve}, we plot the forecasted sensitivity to the characteristic strain $h_c(f)$ for an experiment with $\Delta_T=1\,\mu \mathrm{K}'$, $f_{\text{sky}}=0.5$, $\Delta t_{\text{survey}}=10\,\text{yr}$. The sensitivity curve for our phase-folding method (red) exhibits the characteristic $f^{1/2}$ scaling derived from our frequency-independent noise variance, becoming less sensitive at higher frequencies. We show this in context with other major gravitational wave observatories: pulsar timing arrays (NANOGrav 15-yr \cite{Agazie:2023}), the future space-based interferometer LISA \cite{Colpi:2024}, and ground-based detectors (Advanced LIGO \cite{LIGOScientificCollaboration:2015}). While our method reproduces the sensitivity of time-division techniques \cite{Leluc:2025} at frequencies below $\sim 10^{-5}$\,Hz, its key advantage is overcoming their daily-cadence limitation. The phase-folding framework extends the search for GWs by several orders of magnitude in frequency, opening a continuous window from the time-division floor all the way to the detector's Nyquist limit ($\sim 100$\,Hz). The curve shown here extends only to 20\,Hz, assuming $N_b=10$ phase bins and a 200\,Hz sampling rate; sensitivity at higher frequencies would require fewer phase bins, with $N_b=2$ reaching the Nyquist limit but with a window function factor $W(N_b=2) \approx 0.64$.

While the projected sensitivity is, as expected, significantly less constraining than that of dedicated observatories in their respective bands of peak sensitivity, this work presents a proof-of-concept for a new capability of CMB experiments: probing SGWB over a vast frequency range. The method's primary value lies in its coverage of otherwise difficult-to-measure frequency windows, such as the microhertz-to-hertz regime. In addition, while observatories like LISA are planned for the future, vast archives of high-cadence data from past and current CMB experiments already exist. By performing targeted frequency searches on this data, our technique can be used to place point-by-point constraints on a continuous SGWB or to conduct dedicated searches for coherent periodic signals. This provides a powerful, ready-to-deploy framework for placing the first observational limits in this underexplored territory and ruling out exotic cosmological models that predict an anomalously strong gravitational wave presence.

\section{Time-Dependent Cosmic Birefringence}
\label{sec:biref_app}
Beyond gravitational waves, the phase-folding technique also provides a powerful probe for parity-violating phenomena. A compelling target is cosmic birefringence, a rotation of the linear polarization of CMB photons as they propagate through space. This effect can arise if a pseudoscalar field, such as an ultralight axion-like particle (ALP), couples to photons \cite{Carroll:1990, Harari:1992, Carroll:1998}. A key feature of this mechanism is that the net polarization rotation angle is proportional to the difference in the axion field value at the point of emission and absorption, making it independent of the field's evolution along the photon's path.

If such an ALP constitutes the dark matter, its coherent, oscillating field would induce a time-dependent birefringence angle, $\alpha(t)$, that is uniform across the sky and oscillates at the axion's Compton frequency, $f_a = m_a c^2 / h$ \cite{Fedderke:2019}. This creates a unique, all-sky signal: an oscillation of the CMB polarization pattern at a frequency directly tied to the axion mass. This prospect has motivated a series of targeted searches using time-resolved CMB maps. Recent analyses of time-ordered data from ground-based CMB experiments, including BICEP/Keck \cite{BICEP/Keck:2021, Ade:2022}, POLARBEAR \cite{Adachi:2023}, and SPT-3G \cite{Ferguson:2022}, have searched for this oscillating signal. These studies have placed stringent constraints on the axion-photon coupling $g_{a\gamma\gamma}$ for axion masses in the ``fuzzy dark matter'' range of $m_{a} \sim 10^{-22}-10^{-18}$\,eV, corresponding to oscillation periods from hours to years.

However, these time-division methods are similarly limited by the mapmaking cadence, restricting searches to frequencies below $\sim 10^{-5}$\,Hz, which corresponds to ALP masses $m_a \lesssim 10^{-18}$\,eV. This leaves a vast higher-mass regime largely unexplored by CMB polarization. While other astrophysical probes, such as that from supernovae \cite{Payez:2015} or the solar axion search by the CERN Axion Solar Telescope (CAST) \cite{CASTCollaboration:2017}, set stringent limits, they test axion physics in a fundamentally different regime, sensitive to the relativistic production of axion particles in hot, dense environments. In contrast, a CMB search for polarization oscillations tests the axion as a coherent, non-relativistic classical field that constitutes the local dark matter. This crucial distinction makes the two approaches highly complementary, motivating continued CMB-based searches to probe this unique physical signature at higher mass range.

Our phase-folding framework directly addresses this gap. By enabling targeted searches at frequencies up to the detector's Nyquist limit ($\sim 100$\,Hz), we extend the search range for oscillating cosmic birefringence to ALP masses as high as $m_a \sim 10^{-14}$\,eV.

\subsection{Effects of Cosmic Birefringence on CMB Polarization}
\label{subsec:biref_signal}
A physical rotation of the polarization plane by a time-dependent angle $\alpha(t)$ mixes the intrinsic Stokes parameters, $\tilde{Q}$ and $\tilde{U}$, into the observed ones, $Q$ and $U$
\begin{equation}
\left(Q \pm iU\right)(\hat{\mathbf{n}}, t) = e^{\pm 2i\alpha(t)}\left(\tilde{Q} \pm i\tilde{U}\right)(\hat{\mathbf{n}}).
\label{eq:qu_rotation}
\end{equation}
For each component, this becomes
\begin{equation}
\begin{split}
Q(\hat{\mathbf{n}}, t) &= \tilde{Q}(\hat{\mathbf{n}}) \cos[2\alpha(t)] - \tilde{U}(\hat{\mathbf{n}}) \sin[2\alpha(t)], \\
U(\hat{\mathbf{n}}, t) &= \tilde{U}(\hat{\mathbf{n}}) \cos[2\alpha(t)] + \tilde{Q}(\hat{\mathbf{n}}) \sin[2\alpha(t)].
\end{split}
\label{eq:QU_rotation}
\end{equation}
For cosmological analysis, the polarization field can be decomposed into parity-even E-modes and parity-odd B-modes via a spin-weighted spherical harmonic transform,
\begin{equation}
\left(\tilde{Q} \pm i\tilde{U}\right)(\hat{\mathbf{n}}) = \sum_{\ell m} (\tilde{E}_{\ell m} \pm i\tilde{B}_{\ell m})\, {}_{\pm2}Y_{\ell m}(\hat{\mathbf{n}}),
\end{equation}
where ${}_{\pm2}Y_{\ell m}$ are the spin-2 spherical harmonics. In this work we assume the birefringence angle $\alpha(t)$ is uniform across the sky. Its effect in harmonic space mixes E with B modes,
\begin{align}
E_{\ell m}(t) &= \tilde{E}_{\ell m} \cos[2\alpha(t)] - \tilde{B}_{\ell m} \sin[2\alpha(t)], \\
B_{\ell m}(t) &= \tilde{B}_{\ell m} \cos[2\alpha(t)] + \tilde{E}_{\ell m} \sin[2\alpha(t)].
\end{align}
For small rotation angles ($\alpha(t) \ll 1$), this simplifies to
\begin{align}
E_{\ell m}(t) &\approx \tilde{E}_{\ell m} - 2\alpha(t)\tilde{B}_{\ell m}, \\
B_{\ell m}(t) &\approx \tilde{B}_{\ell m} + 2\alpha(t)\tilde{E}_{\ell m}.
\end{align}
In the standard cosmological model, the primordial B-mode signal is expected to be much smaller than the E-mode signal ($\tilde{C}_\ell^{BB} \ll \tilde{C}_\ell^{EE}$). The term proportional to $\tilde{B}_{\ell m}$ in the expression for $E_{\ell m}(t)$ is therefore second-order small and can be neglected. This leads to the simplified, approximate relations
\begin{align}
E_{\ell m}(t) &\approx \tilde{E}_{\ell m}, \label{eq:E_approx} \\
B_{\ell m}(t) &\approx \tilde{B}_{\ell m} + 2\alpha(t) \tilde{E}_{\ell m}.
\label{eq:biref_EB_mix}
\end{align}
This shows that an oscillating birefringence angle generates a B-mode signal that is spatially proportional to the static, primordial E-mode pattern. For a monochromatic ALP field, the angle oscillates as $\alpha(t) = \alpha_0 \cos(\omega_a t + \delta)$, where $\omega_a = 2\pi f_a$, which is related to the axion mass by $f_a = m_a c^2/h$.

\subsection{Phase-Folding Method for Time-Dependent Birefringence}
\label{subsec:biref_formalism}
We now show that the phase-folding mapmaking framework allows us to probe this oscillating birefringence signal, through a targeted search at a given frequency $f_a$ by folding the full time-ordered data. This process yields $N_b$ phase-binned maps, whose harmonic coefficients we denote as $\{E_{\ell m, b}\}$ and $\{B_{\ell m, b}\}$.

We then apply the DFT across the phase-bin axis to obtain the frequency-mode coefficients
\begin{align}
\mathcal{E}_{\ell m}(k) &= \frac{1}{N_b} \sum_{b=0}^{N_b-1} E_{\ell m,b}\, e^{-i 2\pi b k / N_b}, \\
\mathcal{B}_{\ell m}(k) &= \frac{1}{N_b} \sum_{b=0}^{N_b-1} B_{\ell m,b}\, e^{-i 2\pi b k / N_b}.
\end{align}
As established in Sec.~\ref{subsec:stats}, the static primordial CMB signal resides entirely in the $k=0$ mode. A monochromatic signal at frequency $f_a$ will, by construction, be concentrated in the fundamental oscillating mode, $k=1$.

To see this more explicitly, in a single phase bin $b$, Eq.~\eqref{eq:biref_EB_mix} becomes
\begin{equation}
\begin{split}
E_{\ell m,b} &\approx \tilde{E}_{\ell m},\\
B_{\ell m,b} &\approx \tilde{B}_{\ell m} + 2\alpha_b \tilde{E}_{\ell m},
\end{split}
\label{eq:Bb_linear}
\end{equation}
where $\alpha_b$ is the birefringence angle averaged over the time samples in phase bin $b$. For an oscillation with $\alpha(t) = \alpha_0 \cos(\omega_a t + \delta)$, the value in bin $b$ is
\begin{equation}
\alpha_b = \alpha_0 \cos(\varphi_b + \delta) W(N_b),
\end{equation}
where $\varphi_b$ is the central phase given by $\varphi_b = 2\pi b / N_b$.

The DFT of the binned angle, which we denote $\mathcal{\alpha}_k$, isolates the oscillating component
\begin{equation}
\begin{split}
\mathcal{\alpha}_k &= \frac{1}{N_b} \sum_{b=0}^{N_b-1} \alpha_b\, e^{-i 2\pi b k / N_b} \\
&= \frac{\alpha_0 W(N_b)}{2} \left[ e^{i\delta} \delta_{k,1} + e^{-i\delta} \delta_{k,-1} \right].
\end{split}
\end{equation}
The $k=1$ mode contains the amplitude of the signal,
\begin{equation}
\mathcal{\alpha}_1 = \frac{\alpha_0 W(N_b)}{2} e^{i\delta}.
\label{eq:alpha1_def}
\end{equation}
The $k=-1$ mode is simply its complex conjugate, $\mathcal{\alpha}_{-1} = \mathcal{\alpha}_1^*$, as required for the transform of a real signal, and thus contains no independent information. We therefore focus our analysis on the $k=1$ mode. Substituting Eq.~\eqref{eq:Bb_linear} into the DFT for the B-modes yields
\begin{align}
\mathcal{B}_{\ell m}(k) &= \frac{1}{N_b} \sum_{b=0}^{N_b-1} \left[ \tilde{B}_{\ell m} + 2\alpha_b \tilde{E}_{\ell m} \right] e^{-i 2\pi b k / N_b} \nonumber \\
&= \tilde{B}_{\ell m} \delta_{k,0} + 2\tilde{E}_{\ell m} \mathcal{\alpha}_k.
\end{align}
For the oscillating mode ($k=1$), the primordial B-mode term vanishes, leaving the key signal expression:
\begin{equation}
\mathcal{B}_{\ell m}(k=1) = 2 \mathcal{\alpha}_1 \tilde{E}_{\ell m}.
\label{eq:Bk1_signal}
\end{equation}
This simple result shows that a monochromatic oscillating birefringence signal rotates the E-mode map from $k=0$ into a B-mode map at $k=1$. This causes a non-zero cross-power spectrum between the static E-modes and the oscillating B-modes,
\begin{equation}
\langle \mathcal{E}^*_{\ell m}(0)\, \mathcal{B}_{\ell' m'}(1) \rangle = \langle \tilde{E}^*_{\ell m} \left( 2\mathcal{\alpha}_1 \tilde{E}_{\ell' m'} \right) \rangle = 2\mathcal{\alpha}_1 \tilde{C}_\ell^{EE}\, \delta_{\ell\ell'}\delta_{mm'}.
\label{eq:EB_correlation}
\end{equation}
This suggests that by measuring the cross spectra, $
\hat{C}_\ell^{\mathcal{E}(0)\mathcal{B}(1)}$, one can get an estimate of $\alpha_1$ if a fiducial $\tilde{C}_\ell^{\rm EE}$ is known.

We can then construct a maximum likelihood estimator for $\mathcal{\alpha}_1$ by adapting the formalism used for static birefringence searches \cite{Keating:2013}, with
\begin{equation}
\hat{\mathcal{\alpha}}_1 = \frac{A}{2B},
\label{eq:alpha_estimator}
\end{equation}
where $A$ and $B$ are given by
\begin{align}
A &= \sum_{\ell=\ell_{\text{min}}}^{\ell_{\text{max}}} \frac{2\ell+1}{2} \frac{\hat{C}_\ell^{\mathcal{E}(0)\mathcal{B}(1)} \tilde{C}_\ell^{EE}}{C_\ell^{\mathcal{E},\text{obs}}(0)\, N_\ell^{\mathcal{B}}(1)}, \\
B &= \sum_{\ell=\ell_{\text{min}}}^{\ell_{\text{max}}} \frac{2\ell+1}{2} \frac{(\tilde{C}_\ell^{EE})^2}{C_\ell^{\mathcal{E},\text{obs}}(0)\, N_\ell^{\mathcal{B}}(1)}.
\end{align}
where $C_\ell^{\mathcal{E},\text{obs}}(0) \equiv \tilde{C}_\ell^{EE} + \tilde{C}_\ell^{\rm EE,fg}+ N_\ell^{\mathcal{E}}(0)$ is the total power in the static ($k=0$) E-mode map, including the primordial contribution, $\tilde{C}_\ell^{EE}$, the foreground contribution, $\tilde{C}_\ell^{\rm EE,fg}$, and instrument noise, $N_\ell^{\mathcal{E}}(0)$. $N_\ell^{\mathcal{B}}(1)$ is the instrument noise power in the oscillating ($k=1$) B-mode map, which contains only noise in the absence of oscillating birefringence signal.

This expression is the direct analogue of the estimator for a static birefringence search \cite{Keating:2013}, demonstrating that the mapping from a static operator to a time-dependent one simply requires replacing one leg of the estimator with the corresponding $k=1$ frequency-mode map.

\subsection{A Toy Model}
\label{subsec:biref_toy_model}
\begin{figure}
    \centering
    \includegraphics[width=\linewidth]{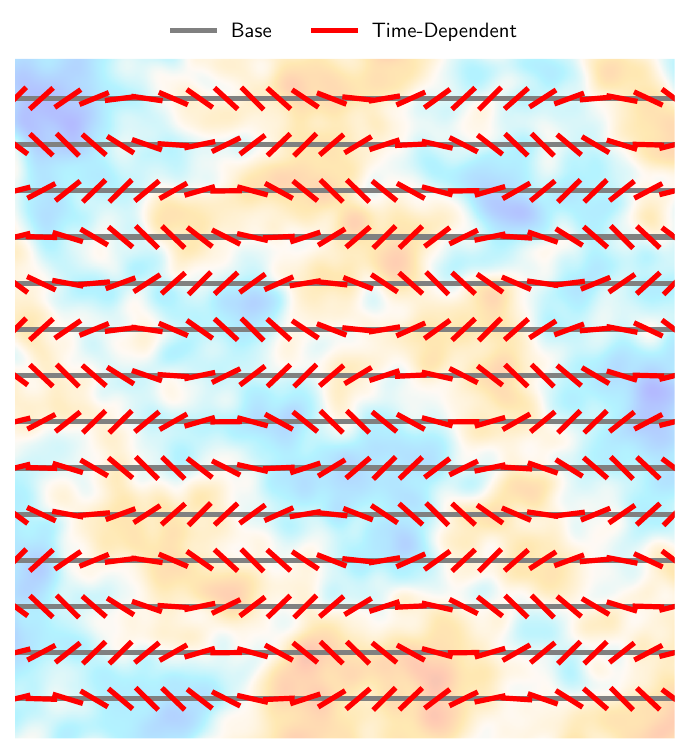}
    \caption{This diagram illustrates of the effect of a 45$^\circ$ time-dependent polarization rotation oscillating at around 1.2\,Hz on the detector polarization response at various points in a scan. The black solid lines show the unrotated detector polarization response, which is a constant on the sky for illustration. The red lines show the effective detector polarization response caused by time-dependent polarization rotations. While the scan covers every pixel in the map, we only show selected paths for illustration. The background shows the primary CMB field.}
    \label{fig:biref_scan_path_illustration}
\end{figure}

To validate the end-to-end formalism, we build a toy model simulation, where we inject a known time-dependent birefringence signal into a detector timestream and then recover the input signal using phase-folded maps.

We begin by generating an unrotated CMB sky realization on a $10^\circ \times 10^\circ$ patch with $0.5'$ pixel resolution. The T, Q, and U maps are simulated with the best-fit \textit{Planck} 2018 power spectra \cite{Planck2018:VI:CP} and are convolved with a Gaussian beam of $\theta_{\rm fwhm} = 1.4'$. We simulate a time-dependent birefringence with amplitude $\alpha_0 = 0.3^\circ$ oscillating at a frequency $f_a = 10$\,Hz. The physical effect of this rotation on the CMB polarization, illustrated in Fig.~\ref{fig:biref_scan_path_illustration}, is implemented at the timestream level. For a detector with instrumental polarization angle $\psi_{\text{inst}}$ observing a sky position $\hat{\mathbf{n}}(t)$ at time $t$, the time-ordered data is modeled as
\begin{equation}
\begin{split}
d(t) &= T(\hat{\mathbf{n}}(t)) + Q(\hat{\mathbf{n}}(t))\cos[2(\psi_{\text{inst}} - \alpha(t))] \\
&\quad + U(\hat{\mathbf{n}}(t))\sin[2(\psi_{\text{inst}} - \alpha(t))] + n(t),
\label{eq:tod_model_biref}
\end{split}
\end{equation}
where $\alpha(t) = \alpha_0 \cos(2\pi f_a t + \delta)$ is the time-dependent rotation angle with initial phase $\delta$, and $n(t)$ is the instrumental noise. This formulation is equivalent to applying the rotation transformation in Eq.~\eqref{eq:qu_rotation} to the sky's Stokes parameters but is computationally more efficient as it avoids regenerating rotated sky maps for each time sample. Instead, we effectively rotate the detector's polarization response by $-\alpha(t)$, which produces the same physical observable.

We simulate a total of 200 single-detector observations to build up sufficient signal-to-noise for signal recovery. To ensure complete coverage of all detector orientations and to avoid degenerate solutions for the three Stokes parameters, we employ four complementary scanning strategies: left-right and up-down boustrophedon (raster) patterns, each repeated with the detector polarization angle rotated by $45^\circ$. These four scan types are cycled through 50 independent realizations, each with a different instrumental polarization angle $\psi_{\text{inst}}$ uniformly sampled from $0$ to $180^\circ$. This sampling strategy ensures that the T, Q, and U components are well-determined at every pixel despite using only single-detector observations. Each scan is sampled at 200\,Hz, providing a Nyquist frequency of 100\,Hz that is well above our target signal frequency of 10\,Hz. We add Gaussian white noise at a level of $4\,\mu\mathrm{K}\sqrt{\text{s}}$ to the time-ordered data, which results in a combined map noise of approximately $2\,\mu\mathrm{K}'$ after inverse-variance weighted coadding of all 200 observations. To realistically model a multi-detector observation where different detectors sample the oscillating signal at different phases, we assign independent random initial phases $\delta$ to the birefringence oscillation for each pair of left-right and up-down scans.

The simulated TODs from all 200 scans are processed through a maximum-likelihood phase-resolved mapmaking algorithm, with $N_b=10$ phase bins. 
This produces a set of 10 phase-binned TQU maps, which collectively capture the full time evolution of the signal over one period of the oscillation. 
We then transform the Q and U maps to E and B using the standard flat-sky pure E/B decomposition in Fourier space. To suppress spectral leakage from the patch edges, we apply a cosine-tapered apodization mask with a width of $0.2^\circ$ to each map before the Fourier transform. We then transform phase-bin maps into frequency maps following Eq.~\eqref{eq:dft_maps} and obtain $\{\mathcal{T}(\hat{\mathbf{n}}, k), \mathcal{E}(\hat{\mathbf{n}}, k), \mathcal{B}(\hat{\mathbf{n}}, k)\}$ for $k = 0, 1, \ldots, N_b-1$.

We then compute the $EB$ cross-spectrum between the static E-mode map ($\mathcal{E}(0)$) and the oscillating B-mode map ($\mathcal{B}(1)$), denoted $\hat{C}_\ell^{\mathcal{E}(0)\mathcal{B}(1)}$, which is expected to contain the birefringence signal according to Eq.~\eqref{eq:EB_correlation}. This spectrum is estimated using a simple FFT-based method with bins of width $\Delta l = 50$ over the multipole range $100 \leq l \leq 3000$.

\begin{figure}
    \centering
    \includegraphics[width=\linewidth]{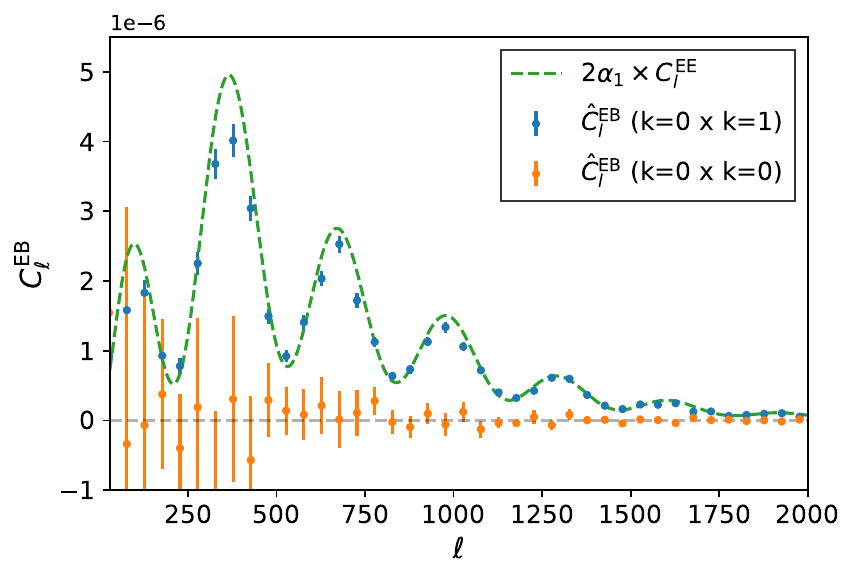}
    \caption{Demonstration of birefringence signal recovery from the toy model. The blue points show the measured EB cross-spectrum between the static ($k=0$) E-mode and oscillating ($k=1$) B-mode maps. The green dashed line shows the theoretical expectation for the input signal ($2\mathcal{\alpha}_1 C_l^{EE}$), showing a good agreement. In comparison, the orange points show the auto-spectrum of the static ($k=0$) map which is consistent with null, as expected.}
    \label{fig:biref_recovery}
\end{figure}

Figure~\ref{fig:biref_recovery} plots the resulting cross-power spectrum, $\hat{C}_\ell^{\mathcal{E}(0)\mathcal{B}(1)}$ (blue points), comparing it to the theoretical expectation from the input signal, $2\mathcal{\alpha}_{1}C_{\ell}^{\tilde{E}\tilde{E}}$ (green dashed line). The measured spectrum demonstrates good agreement with the theory. We note a small suppression of power at large scales, which is likely a result of the sky patch geometry and the associated window function---effects we have neglected in this simple demonstration but which must be considered in a realistic measurement. For comparison, we also show the cross-power spectrum of the static maps, $\hat{C}_\ell^{\mathcal{E}(0)\mathcal{B}(0)}$ (orange points), which is consistent with null, as expected. This result confirms that while a time-dependent polarization rotation is undetectable in the time-averaged map, it is successfully recovered through the coupling between the phase-folded $k=0$ and $k=1$ modes.

Using this measured cross-spectrum as input for the maximum likelihood estimator in Eq.~\eqref{eq:alpha_estimator}, we compute the final signal amplitude. Applying the estimator over the multipole range $30 \leq l \leq 3000$, we recover an oscillation amplitude of $\alpha_0 = 0.27 \pm 0.04^\circ$. This recovered value is in good agreement with the input amplitude of $0.3^\circ$ within $1\sigma$. This successful recovery validates the entire phase-folding and estimation framework as a powerful method for isolating a time-dependent birefringence signature from the static CMB signal.

\subsection{Sensitivity Forecast}
\label{subsec:biref_sensitivity}
\begin{figure}
  \centering
  \includegraphics[width=\linewidth]{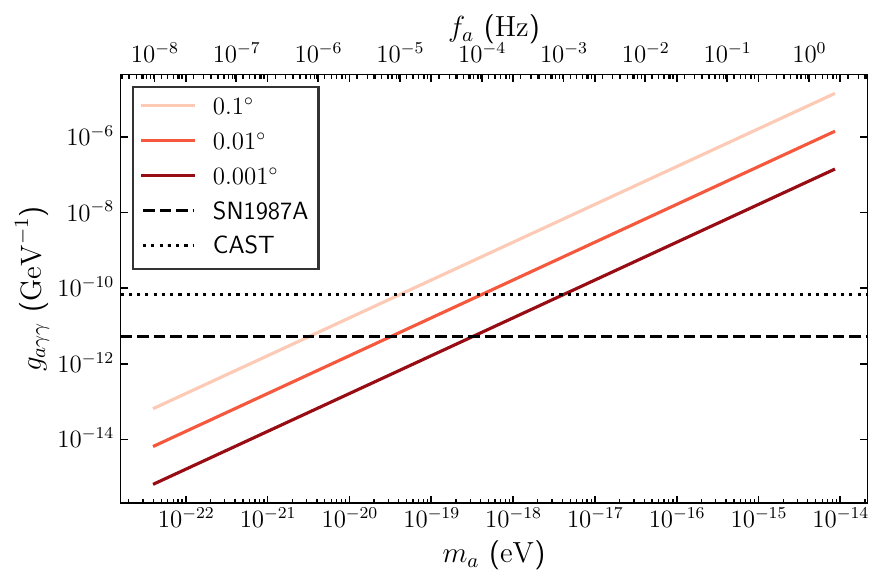}
  \caption{Projected constraints on the axion-photon coupling constant, $g_{a\gamma\gamma}$, as a function of axion mass, $m_a$, achievable with the phase-folding method. The solid curves show the forecasted upper limits for different experimental sensitivities to the oscillating birefringence amplitude, $\sigma(\alpha_0) = 0.1^\circ$, $0.01^\circ$, and $0.001^\circ$ (from lightest to darkest). This forecast assumes the local axion dark matter density is $\rho_a = 0.3\,\mathrm{GeV/cm}^{3}$. For context, we compare our projected reach with existing upper limits from the supernova SN1987A (dashed line) \cite{Payez:2015} and the CERN Axion Solar Telescope (CAST) experiment (dotted line) \cite{CASTCollaboration:2017}. The phase-folding technique enables CMB experiments to probe a new region of parameter space at high axion masses, complementary to existing laboratory and astrophysical searches.}
  \label{fig:g_ag_sensitivity}
\end{figure}

Having validated the estimator with our toy model, we now forecast the sensitivity of this method to oscillating cosmic birefringence as a function of the axion frequency $f_a$. The variance of the estimator $\hat{\mathcal{\alpha}}_1$ defines the fundamental sensitivity of our method, which is given by
\begin{equation}
  \sigma(\alpha_1)^2 = \left[ 2 f_{\text{sky}} \sum_{\ell=\ell_{\text{min}}}^{\ell_{\text{max}}} (2\ell+1) \frac{(\tilde{C}_\ell^{EE})^2}{C_\ell^{\mathcal{E},\text{obs}}(0)\, N_\ell^{\mathcal{B}}(1)} \right]^{-1},
  \label{eq:sigma_a1}
\end{equation}
where $C_\ell^{\mathcal{E},\text{obs}}(0)$ is the total power in the static E-mode map, and $N_\ell^{\mathcal{B}}(1) = N_\ell^{\mathcal{B},\text{total}}$ is the instrumental noise power in the oscillating B-mode map. Note that the dominant source of B-modes, CMB lensing, is a static effect and therefore does not contribute to the variance in the $k=1$ mode.

The physical amplitude of the oscillation is $\alpha_0$, related to the complex Fourier amplitude $\mathcal{\alpha}_1$ by Eq.~\eqref{eq:alpha1_def}. Its variance is given by
\begin{equation}
\sigma(\alpha_0) = \frac{\sqrt{2}}{W(N_b)} \, \sigma(\mathcal{\alpha}_1).
\label{eq:alpha0_sensitivity}
\end{equation}
where $W(N_b)$ is the window function, and the factor $\sqrt{2}$ accounts for the fact that $\mathcal{\alpha}_1$ is a complex quantity whose real and imaginary parts are independently measured. It's worth noting that the sensitivity $\sigma(\alpha_0)$ is independent of the target frequency $f_a$. For a large-aperture CMB experiment with $\ell_{\rm min}=30$ and $\ell_{\rm max}=3000$, we find
\begin{equation}
\sigma(\alpha_0) \approx 1.6 \times10^{-3} \,\mathrm{deg} \left[\frac{\Delta_T}{1\,\mu\mathrm{K}'}\right]\left[ \frac{f_{\rm sky}}{0.5} \right]^{-\frac{1}{2}}.
\label{eq:alpha0_sensitivity_forecast}
\end{equation}

We can translate the sensitivity on the rotation angle $\alpha_0$ into a constraint on the axion-photon coupling constant $g_{a\gamma\gamma}$. If an ultralight axion-like particle constitutes the local dark matter density $\rho_0$, its field oscillates as $\phi(t) = \phi_0 \cos(m_a t)$, where \(m_a\) is the axion mass and the amplitude is \(\phi_0 = \sqrt{2\rho_0}/m_a\) \cite{Marsh:2016:axion}. This oscillating field induces a time-dependent birefringence angle
\begin{equation}
    \alpha(t) = \frac{g_{a\gamma\gamma}}{2} \phi(t) = \frac{g_{a\gamma\gamma} \sqrt{2\rho_0}}{2 m_a} \cos(m_a t + \delta).
\end{equation}
Comparing this to our signal model \(\alpha(t) = \alpha_0 \cos(\omega_a t + \delta)\), we identify the amplitude as
\begin{equation}
\alpha_0 = \frac{g_{a\gamma\gamma} \sqrt{\rho_0}}{\sqrt{2} \, m_a}.
\end{equation}
which yield
$g_{a\gamma\gamma} = \sqrt{2} \alpha_0 m_a/\sqrt{\rho_0}$. This shows that, for a fixed observable rotation $\alpha_0$, the required coupling $g_{a\gamma\gamma}$ scales linearly with the axion mass $m_a$. The uncertainty on the coupling constant is therefore
\begin{equation}
  \begin{split}
    \sigma(g_{a\gamma\gamma}) &= \frac{\sqrt{2} \, m_a}{\sqrt{\rho_0}} \, \sigma(\alpha_0) \\ &\approx 1.63\times10^{-14} \,\mathrm{GeV}^{-1} \left[\frac{m_a}{10^{-14}\,\mathrm{eV}} \right] \left[\frac{\sigma(\alpha_0)}{0.01^\circ}\right]  \\
    &\qquad \times \left[\frac{\kappa \rho_0}{0.3\,\mathrm{GeV/cm}^{3}}\right]^{-\frac{1}{2}}.
  \end{split}
  \label{eq:sigma_g}
\end{equation}
where $\sigma(\alpha_{0})$ is given by Eq.~\eqref{eq:alpha0_sensitivity_forecast}, $\rho_0$ represents the local dark matter density, $\kappa$ represents the fraction of dark matter density constituted by axion dark matter. This result encapsulates the key trade-off in our sensitivity forecast. While the estimator's sensitivity to the rotation angle \(\sigma(\alpha_0)\) is frequency-independent, the sensitivity to the fundamental coupling constant \(\sigma(g_{a\gamma\gamma})\) scales linearly with mass \(m_a\). This scaling arises from the fact that, for a fixed local dark matter density, a higher-mass axion has a smaller field amplitude \(\phi_0\), and therefore requires a stronger coupling \(g_{a\gamma\gamma}\) to produce a detectable rotation \(\alpha_0\).

Fig.~\ref{fig:g_ag_sensitivity} illustrates the projected constraints on the axion-photon coupling \(g_{a\gamma\gamma}\) for different levels of experimental sensitivity \(\sigma(\alpha_0)\). Our method significantly expands the reach of time-division map-based searches for axion dark matter, pushing the accessible mass range from $m_a \lesssim 10^{-18}$ eV (e.g., \cite{Adachi:2023}) up to the $m_a \sim 10^{-14}$ eV scale with $N_b=10$. The projected sensitivity to the coupling constant $g_{a\gamma\gamma}$ degrades linearly with axion mass, as dictated by Eq.~\eqref{eq:sigma_g}, which explains the characteristic upward slope of the forecast curves in the figure.

While our projected constraints in this higher-mass regime are significantly less constraining than existing limits from astrophysical observations from supernova (SN1987A; \cite{Payez:2015}) or laboratory experiments like CAST \cite{CASTCollaboration:2017}, it is crucial to recognize that these probes test axion physics in fundamentally different regimes. CMB-based searches probe coherent, non-relativistic axion field behaving as a classical wave that constitutes the local dark matter. In contrast, the SN1987A and CAST limits are derived from the thermal production of relativistic axion particles in hot, dense environments. These are therefore highly complementary tests of the same fundamental coupling under different physical conditions. Thus, the phase-folding technique provides a powerful new way to place the first direct observational limits on non-relativistic axion dark matter in this largely unexplored high-frequency window.

\section{Conclusion}
\label{sec:conclusion}
We have introduced a framework for extracting frequency-mode signal from CMB observations that significantly extends the accessible frequency range for searches of periodic cosmological and astrophysical phenomena. By adapting phase-folding techniques to CMB mapmaking, we have shown how to construct frequency-mode maps that cleanly separate static sky signals from oscillating components. This approach overcomes the fundamental cadence limitation of conventional time-division methods, pushing the frequency reach from $\sim 10^{-5}$\,Hz to the detector sampling rate of $\mathcal{O}(100)$\,Hz—an improvement of seven orders of magnitude.

We demonstrated the framework's power with two cosmological applications. First, we developed a quadratic estimator (Sec.~\ref{sec:gw_app}) for the time-dependent lensing deflection from a stochastic gravitational wave background (SGWB). This extends CMB-based GW searches into the largely unexplored millihertz-to-hertz regime. While less sensitive than dedicated interferometers, our method uniquely probes challenging frequency windows and can be applied to existing archival data. Second, we formulated an estimator for oscillating cosmic birefringence (Sec.~\ref{sec:biref_app}), enabling searches for ultralight axion-like particles (ALPs) at masses up to $m_a \sim 10^{-14}$\,eV. This extends the mass reach of time-division map-based searches by over four orders of magnitude, probing the axion-photon coupling ($g_{a\gamma\gamma}$) for ALPs behaving as classical, local dark matter. This is complementary to constraints from astrophysical and laboratory searches, which probe relativistic axion particle production.

The phase-folding framework is not limited to these two applications. Any physical process that modulates the CMB signal periodically—whether through lensing, rotation, or other mechanisms—can be targeted with this method, opening avenues for a wide range of searches for time-varying phenomena. This unlocks the scientific potential of vast archival and future high-cadence data from experiments like ACT, SPT, and the Simons Observatory, effectively turning modern CMB surveys into versatile spectrometers for new physics.

A robust application to real data will require careful mitigation of potential systematic effects, such as detector-specific periodic noise and scan-synchronous signals. These can be addressed through cross-detector null tests and instrument modeling. A detailed characterization of the framework's performance in the presence of correlated noise (e.g., atmospheric $1/f$ noise) will be essential, which we leave for future work. With these in place, our framework provides a powerful new tool for extracting time-dependent physics from CMB observations, with immediate applicability to existing and upcoming CMB datasets.

\vspace{0.5cm}
\section*{Acknowledgments}
We thank Adam Hincks for helpful comments. YG acknowledges support from the University of Toronto's Eric and Wendy Schmidt AI in Science Postdoctoral Fellowship, a program of Schmidt Sciences. The Dunlap Institute is funded through an endowment established by the David Dunlap family and the University of Toronto.

\bibliography{cite}

@ARTICLE{ACT:Biermann:2025,
	author = {{Biermann}, Emily and {Li}, Yaqiong and {Naess}, Sigurd and others},
	title = "{The Atacama Cosmology Telescope: Systematic Transient Search of Single Observation Maps}",
	journal = {\apj},
	year = 2025,
	month = jun,
	volume = {986},
	number = {1},
	eid = {7},
	pages = {7},
	doi = {10.3847/1538-4357/adce70},
	archivePrefix = {arXiv},
	eprint = {2409.08429},
	primaryClass = {astro-ph.HE},
	adsurl = {https://ui.adsabs.harvard.edu/abs/2025ApJ...986....7B}
}

@ARTICLE{ACT:Li:2023,
	author = {{Li}, Yaqiong and {Biermann}, Emily and {Naess}, Sigurd and others},
	title = "{The Atacama Cosmology Telescope: Systematic Transient Search of 3 Day Maps}",
	journal = {\apj},
	year = 2023,
	month = oct,
	volume = {956},
	number = {1},
	eid = {36},
	pages = {36},
	doi = {10.3847/1538-4357/ace599},
	archivePrefix = {arXiv},
	eprint = {2303.04767},
	primaryClass = {astro-ph.SR},
	adsurl = {https://ui.adsabs.harvard.edu/abs/2023ApJ...956...36L}
}

@ARTICLE{ACT:Naess:2021,
	author = {{Naess}, Sigurd and others},
	title = "{The Atacama Cosmology Telescope: Detection of Millimeter-wave Transient Sources}",
	journal = {\apj},
	year = 2021,
	month = jul,
	volume = {915},
	number = {1},
	eid = {14},
	pages = {14},
	doi = {10.3847/1538-4357/abfe6d},
	archivePrefix = {arXiv},
	eprint = {2012.14347},
	primaryClass = {astro-ph.SR},
	adsurl = {https://ui.adsabs.harvard.edu/abs/2021ApJ...915...14N}
}

@ARTICLE{ACT:Naess:2021:P9,
	author = {{Naess}, Sigurd and others},
	title = "{The Atacama Cosmology Telescope: A Search for Planet 9}",
	journal = {\apj},
	year = 2021,
	month = dec,
	volume = {923},
	number = {2},
	eid = {224},
	pages = {224},
	doi = {10.3847/1538-4357/ac2307},
	archivePrefix = {arXiv},
	eprint = {2104.10264},
	primaryClass = {astro-ph.EP},
	adsurl = {https://ui.adsabs.harvard.edu/abs/2021ApJ...923..224N}
}

@ARTICLE{ACT:Naess:2025,
	author = {{Naess}, Sigurd and {Guan}, Yilun and {Duivenvoorden}, Adriaan J. and {Hasselfield}, Matthew and {Wang}, Yuhan and others},
	title = "{The Atacama Cosmology Telescope: DR6 Maps}",
	journal = {arXiv e-prints},
	year = 2025,
	month = mar,
	eid = {arXiv:2503.14451},
	pages = {arXiv:2503.14451},
	doi = {10.48550/arXiv.2503.14451},
	archivePrefix = {arXiv},
	eprint = {2503.14451},
	primaryClass = {astro-ph.CO},
	adsurl = {https://ui.adsabs.harvard.edu/abs/2025arXiv250314451N}
}

@ARTICLE{ACT:Orlowski-Scherer:2024,
	author = {{Orlowski-Scherer}, John and {Venterea}, Ricco C. and {Battaglia}, Nicholas and {Naess}, Sigurd and others},
	title = "{The Atacama Cosmology Telescope: Millimeter Observations of a Population of Asteroids or: ACTeroids}",
	journal = {\apj},
	year = 2024,
	month = apr,
	volume = {964},
	number = {2},
	eid = {138},
	pages = {138},
	doi = {10.3847/1538-4357/ad21fe},
	archivePrefix = {arXiv},
	eprint = {2306.05468},
	primaryClass = {astro-ph.EP},
	adsurl = {https://ui.adsabs.harvard.edu/abs/2024ApJ...964..138O}
}

@ARTICLE{Acernese:2015,
	author = {{Acernese}, F. and {Agathos}, M. and {Agatsuma}, K. and others},
	title = "{Advanced Virgo: a second-generation interferometric gravitational wave detector}",
	journal = {Classical and Quantum Gravity},
	year = 2015,
	month = jan,
	volume = {32},
	number = {2},
	eid = {024001},
	pages = {024001},
	doi = {10.1088/0264-9381/32/2/024001},
	archivePrefix = {arXiv},
	eprint = {1408.3978},
	primaryClass = {gr-qc},
	adsurl = {https://ui.adsabs.harvard.edu/abs/2015CQGra..32b4001A}
}

@ARTICLE{Adachi:2023,
	author = {{Polarbear Collaboration} and others},
	title = "{Constraints on axionlike polarization oscillations in the cosmic microwave background with POLARBEAR}",
	journal = {\prd},
	year = 2023,
	month = aug,
	volume = {108},
	number = {4},
	eid = {043017},
	pages = {043017},
	doi = {10.1103/PhysRevD.108.043017},
	archivePrefix = {arXiv},
	eprint = {2303.08410},
	primaryClass = {astro-ph.CO},
	adsurl = {https://ui.adsabs.harvard.edu/abs/2023PhRvD.108d3017A}
}

@ARTICLE{Ade:2022,
	author = {{Bicep/Keck Collaboration} and others},
	title = "{BICEP/Keck XIV: Improved constraints on axionlike polarization oscillations in the cosmic microwave background}",
	journal = {\prd},
	year = 2022,
	month = jan,
	volume = {105},
	number = {2},
	eid = {022006},
	pages = {022006},
	doi = {10.1103/PhysRevD.105.022006},
	archivePrefix = {arXiv},
	eprint = {2108.03316},
	primaryClass = {astro-ph.CO},
	adsurl = {https://ui.adsabs.harvard.edu/abs/2022PhRvD.105b2006A}
}

@ARTICLE{Agazie:2023,
	author = {{Nanograv Collaboration} and others},
	title = "{The NANOGrav 15 yr Data Set: Constraints on Supermassive Black Hole Binaries from the Gravitational-wave Background}",
	journal = {\apjl},
	year = 2023,
	month = aug,
	volume = {952},
	number = {2},
	eid = {L37},
	pages = {L37},
	doi = {10.3847/2041-8213/ace18b},
	archivePrefix = {arXiv},
	eprint = {2306.16220},
	primaryClass = {astro-ph.HE},
	adsurl = {https://ui.adsabs.harvard.edu/abs/2023ApJ...952L..37A}
}

@ARTICLE{BICEP/Keck:2021,
	author = {{BICEP/Keck} and others},
	title = "{BICEP/Keck XII: Constraints on axionlike polarization oscillations in the cosmic microwave background}",
	journal = {\prd},
	year = 2021,
	month = feb,
	volume = {103},
	number = {4},
	eid = {042002},
	pages = {042002},
	doi = {10.1103/PhysRevD.103.042002},
	archivePrefix = {arXiv},
	eprint = {2011.03483},
	primaryClass = {astro-ph.CO},
	adsurl = {https://ui.adsabs.harvard.edu/abs/2021PhRvD.103d2002B}
}

@ARTICLE{Book:2011,
	author = {{Book}, Laura G. and {Flanagan}, {\'E}anna {\'E}.},
	title = "{Astrometric effects of a stochastic gravitational wave background}",
	journal = {\prd},
	year = 2011,
	month = jan,
	volume = {83},
	number = {2},
	eid = {024024},
	pages = {024024},
	doi = {10.1103/PhysRevD.83.024024},
	archivePrefix = {arXiv},
	eprint = {1009.4192},
	primaryClass = {astro-ph.CO},
	adsurl = {https://ui.adsabs.harvard.edu/abs/2011PhRvD..83b4024B}
}

@ARTICLE{CASTCollaboration:2017,
	author = {{CAST Collaboration} and others},
	title = "{New CAST limit on the axion{\textendash}photon interaction}",
	journal = {Nature Physics},
	year = 2017,
	month = jun,
	volume = {13},
	number = {6},
	pages = {584-590},
	doi = {10.1038/nphys4109},
	archivePrefix = {arXiv},
	eprint = {1705.02290},
	primaryClass = {hep-ex},
	adsurl = {https://ui.adsabs.harvard.edu/abs/2017NatPh..13..584C}
}

@ARTICLE{Carroll:1990,
	author = {{Carroll}, Sean M. and {Field}, George B. and {Jackiw}, Roman},
	title = "{Limits on a Lorentz- and parity-violating modification of electrodynamics}",
	journal = {\prd},
	year = 1990,
	month = feb,
	volume = {41},
	number = {4},
	pages = {1231-1240},
	doi = {10.1103/PhysRevD.41.1231},
	adsurl = {https://ui.adsabs.harvard.edu/abs/1990PhRvD..41.1231C}
}

@article{Carroll:1998,
	title = {Quintessence and the {{Rest}} of the {{World}}: {{Suppressing Long-Range Interactions}}},
	author = {Carroll, Sean M},
	year = {1998},
	month = oct,
	journal = {{\textbackslash}prl},
	volume = {81},
	number = {15},
	eprint = {astro-ph/9806099},
	pages = {3067--3070},
	doi = {10.1103/PhysRevLett.81.3067},
	archiveprefix = {arXiv}
}

@ARTICLE{Chown:2018,
	author = {{Chown}, R. and {Omori}, Y. and others},
	title = "{Maps of the Southern Millimeter-wave Sky from Combined 2500 deg$^{2}$ SPT-SZ and Planck Temperature Data}",
	journal = {\apjs},
	year = 2018,
	month = nov,
	volume = {239},
	number = {1},
	eid = {10},
	pages = {10},
	doi = {10.3847/1538-4365/aae694},
	archivePrefix = {arXiv},
	eprint = {1803.10682},
	primaryClass = {astro-ph.CO},
	adsurl = {https://ui.adsabs.harvard.edu/abs/2018ApJS..239...10C}
}

@ARTICLE{Colpi:2024,
	author = {{Colpi}, Monica and {Danzmann}, Karsten and {Hewitson}, Martin and others},
	title = "{LISA Definition Study Report}",
	journal = {arXiv e-prints},
	year = 2024,
	month = feb,
	eid = {arXiv:2402.07571},
	pages = {arXiv:2402.07571},
	doi = {10.48550/arXiv.2402.07571},
	archivePrefix = {arXiv},
	eprint = {2402.07571},
	primaryClass = {astro-ph.CO},
	adsurl = {https://ui.adsabs.harvard.edu/abs/2024arXiv240207571C}
}

@ARTICLE{Darling:2018,
	author = {{Darling}, Jeremy and {Truebenbach}, Alexandra E. and {Paine}, Jennie},
	title = "{Astrometric Limits on the Stochastic Gravitational Wave Background}",
	journal = {\apj},
	year = 2018,
	month = jul,
	volume = {861},
	number = {2},
	eid = {113},
	pages = {113},
	doi = {10.3847/1538-4357/aac772},
	archivePrefix = {arXiv},
	eprint = {1804.06986},
	primaryClass = {astro-ph.IM},
	adsurl = {https://ui.adsabs.harvard.edu/abs/2018ApJ...861..113D}
}

@ARTICLE{Fedderke:2019,
	author = {{Fedderke}, Michael A. and {Graham}, Peter W. and {Rajendran}, Surjeet},
	title = "{Axion dark matter detection with CMB polarization}",
	journal = {\prd},
	year = 2019,
	month = jul,
	volume = {100},
	number = {1},
	eid = {015040},
	pages = {015040},
	doi = {10.1103/PhysRevD.100.015040},
	archivePrefix = {arXiv},
	eprint = {1903.02666},
	primaryClass = {astro-ph.CO},
	adsurl = {https://ui.adsabs.harvard.edu/abs/2019PhRvD.100a5040F}
}

@ARTICLE{Ferguson:2022,
	author = {{Ferguson}, K.~R. and {Anderson}, A.~J. and {Whitehorn}, N. and others},
	title = "{Searching for axionlike time-dependent cosmic birefringence with data from SPT-3G}",
	journal = {\prd},
	year = 2022,
	month = aug,
	volume = {106},
	number = {4},
	eid = {042011},
	pages = {042011},
	doi = {10.1103/PhysRevD.106.042011},
	archivePrefix = {arXiv},
	eprint = {2203.16567},
	primaryClass = {astro-ph.CO},
	adsurl = {https://ui.adsabs.harvard.edu/abs/2022PhRvD.106d2011F}
}

@ARTICLE{GaiaCollaboration:2016,
	author = {{Gaia Collaboration} and others},
	title = "{The Gaia mission}",
	journal = {\aap},
	year = 2016,
	month = nov,
	volume = {595},
	eid = {A1},
	pages = {A1},
	doi = {10.1051/0004-6361/201629272},
	archivePrefix = {arXiv},
	eprint = {1609.04153},
	primaryClass = {astro-ph.IM},
	adsurl = {https://ui.adsabs.harvard.edu/abs/2016A&A...595A...1G}
}

@ARTICLE{Guan:2025,
	author = {{Guan}, Yilun and {Naess}, Sigurd and {Niebres}, Ian and {Branch}, Louis and {Hincks}, Adam D. and others},
	title = "{Atacama Cosmology Telescope: Constraints on the Millimetre Flux of the Crab Pulsar}",
	journal = {arXiv e-prints},
	year = 2025,
	month = sep,
	eid = {arXiv:2509.11960},
	pages = {arXiv:2509.11960},
	doi = {10.48550/arXiv.2509.11960},
	archivePrefix = {arXiv},
	eprint = {2509.11960},
	primaryClass = {astro-ph.HE},
	adsurl = {https://ui.adsabs.harvard.edu/abs/2025arXiv250911960G}
}

@ARTICLE{Harari:1992,
	author = {{Harari}, Diego and {Sikivie}, Pierre},
	title = "{Effects of a Nambu-Goldstone boson on the polarization of radio galaxies and the cosmic microwave background}",
	journal = {Physics Letters B},
	year = 1992,
	month = sep,
	volume = {289},
	number = {1-2},
	pages = {67-72},
	doi = {10.1016/0370-2693(92)91363-E},
	adsurl = {https://ui.adsabs.harvard.edu/abs/1992PhLB..289...67H}
}

@ARTICLE{Jaraba:2023,
	author = {{Jaraba}, Santiago and {Garc{\'\i}a-Bellido}, Juan and {Kuroyanagi}, Sachiko and {Ferraiuolo}, Sarah and {Braglia}, Matteo},
	title = "{Stochastic gravitational wave background constraints from Gaia DR3 astrometry}",
	journal = {\mnras},
	year = 2023,
	month = sep,
	volume = {524},
	number = {3},
	pages = {3609-3622},
	doi = {10.1093/mnras/stad2141},
	archivePrefix = {arXiv},
	eprint = {2304.06350},
	primaryClass = {astro-ph.CO},
	adsurl = {https://ui.adsabs.harvard.edu/abs/2023MNRAS.524.3609J}
}

@article{KKS:1997,
	archivePrefix = {arXiv},
	arxivId = {astro-ph/9611125},
	author = {Kamionkowski, Marc and Kosowsky, Arthur and Stebbins, Albert},
	doi = {10.1103/PhysRevD.55.7368},
	eprint = {9611125},
	issn = {15502368},
	journal = {Physical Review D - Particles, Fields, Gravitation and Cosmology},
	number = {12},
	pages = {7368--7388},
	primaryClass = {astro-ph},
	title = {{Statistics of cosmic microwave background polarization}},
	volume = {55},
	year = {1997}
}

@ARTICLE{Keating:2013,
	author = {{Keating}, Brian G. and {Shimon}, Meir and {Yadav}, Amit P.~S.},
	title = "{Self-calibration of Cosmic Microwave Background Polarization Experiments}",
	journal = {\apjl},
	year = 2013,
	month = jan,
	volume = {762},
	number = {2},
	eid = {L23},
	pages = {L23},
	doi = {10.1088/2041-8205/762/2/L23},
	archivePrefix = {arXiv},
	eprint = {1211.5734},
	primaryClass = {astro-ph.CO},
	adsurl = {https://ui.adsabs.harvard.edu/abs/2013ApJ...762L..23K}
}

@ARTICLE{LIGOScientificCollaboration:2015,
	author = {{LIGO Scientific Collaboration} and others},
	title = "{Advanced LIGO}",
	journal = {Classical and Quantum Gravity},
	year = 2015,
	month = apr,
	volume = {32},
	number = {7},
	eid = {074001},
	pages = {074001},
	doi = {10.1088/0264-9381/32/7/074001},
	archivePrefix = {arXiv},
	eprint = {1411.4547},
	primaryClass = {gr-qc},
	adsurl = {https://ui.adsabs.harvard.edu/abs/2015CQGra..32g4001L}
}

@ARTICLE{Leluc:2025,
	author = {{Leluc}, Alvin and {Meyers}, Joel and {van Engelen}, Alexander},
	title = "{Time-Dependent Deflection Reconstruction: A New Technique to Search for Gravitational Waves with the Cosmic Microwave Background}",
	journal = {arXiv e-prints},
	year = 2025,
	month = aug,
	eid = {arXiv:2508.10673},
	pages = {arXiv:2508.10673},
	doi = {10.48550/arXiv.2508.10673},
	archivePrefix = {arXiv},
	eprint = {2508.10673},
	primaryClass = {astro-ph.CO},
	adsurl = {https://ui.adsabs.harvard.edu/abs/2025arXiv250810673L}
}

@ARTICLE{Marsh:2016:axion,
	author = {{Marsh}, David J.~E.},
	title = "{Axion cosmology}",
	journal = {\physrep},
	year = 2016,
	month = jul,
	volume = {643},
	pages = {1-79},
	doi = {10.1016/j.physrep.2016.06.005},
	archivePrefix = {arXiv},
	eprint = {1510.07633},
	primaryClass = {astro-ph.CO},
	adsurl = {https://ui.adsabs.harvard.edu/abs/2016PhR...643....1M}
}

@ARTICLE{Payez:2015,
	author = {{Payez}, Alexandre and {Evoli}, Carmelo and {Fischer}, Tobias and {Giannotti}, Maurizio and {Mirizzi}, Alessandro and {Ringwald}, Andreas},
	title = "{Revisiting the SN1987A gamma-ray limit on ultralight axion-like particles}",
	journal = {\jcap},
	year = 2015,
	month = feb,
	volume = {2015},
	number = {2},
	pages = {006-006},
	doi = {10.1088/1475-7516/2015/02/006},
	archivePrefix = {arXiv},
	eprint = {1410.3747},
	primaryClass = {astro-ph.HE},
	adsurl = {https://ui.adsabs.harvard.edu/abs/2015JCAP...02..006P}
}

@ARTICLE{Planck2018:VI:CP,
	author = {{Planck Collaboration} and others},
	title = "{Planck 2018 results. VI. Cosmological parameters}",
	journal = {\aap},
	year = 2020,
	month = sep,
	volume = {641},
	eid = {A6},
	pages = {A6},
	doi = {10.1051/0004-6361/201833910},
	archivePrefix = {arXiv},
	eprint = {1807.06209},
	primaryClass = {astro-ph.CO},
	adsurl = {https://ui.adsabs.harvard.edu/abs/2020A&A...641A...6P}
}

@ARTICLE{SPT:Chichura:2022,
	author = {{Chichura}, P.~M. and {Foster}, A. and {Patel}, C. and {Ossa-Jaen}, N. and others},
	title = "{Asteroid Measurements at Millimeter Wavelengths with the South Pole Telescope}",
	journal = {\apj},
	year = 2022,
	month = sep,
	volume = {936},
	number = {2},
	eid = {173},
	pages = {173},
	doi = {10.3847/1538-4357/ac89ec},
	archivePrefix = {arXiv},
	eprint = {2202.01406},
	primaryClass = {astro-ph.EP},
	adsurl = {https://ui.adsabs.harvard.edu/abs/2022ApJ...936..173C}
}

@ARTICLE{SPT:Guns:2021,
	author = {{Guns}, S. and {Foster}, A. and {Daley}, C. and {Rahlin}, A. and {Whitehorn}, N. and others},
	title = "{Detection of Galactic and Extragalactic Millimeter-wavelength Transient Sources with SPT-3G}",
	journal = {\apj},
	year = 2021,
	month = aug,
	volume = {916},
	number = {2},
	eid = {98},
	pages = {98},
	doi = {10.3847/1538-4357/ac06a3},
	archivePrefix = {arXiv},
	eprint = {2103.06166},
	primaryClass = {astro-ph.HE},
	adsurl = {https://ui.adsabs.harvard.edu/abs/2021ApJ...916...98G}
}

@ARTICLE{SPT:Whitehorn:2016,
	author = {{Whitehorn}, N. and {Natoli}, T. and others},
	title = "{Millimeter Transient Point Sources in the SPTpol 100 Square Degree Survey}",
	journal = {\apj},
	year = 2016,
	month = oct,
	volume = {830},
	number = {2},
	eid = {143},
	pages = {143},
	doi = {10.3847/0004-637X/830/2/143},
	archivePrefix = {arXiv},
	eprint = {1604.03507},
	primaryClass = {astro-ph.HE},
	adsurl = {https://ui.adsabs.harvard.edu/abs/2016ApJ...830..143W}
}

@ARTICLE{Zaldarriaga:1997,
	author = {{Zaldarriaga}, Matias and {Seljak}, Uro{\v{s}}},
	title = "{All-sky analysis of polarization in the microwave background}",
	journal = {\prd},
	year = 1997,
	month = feb,
	volume = {55},
	number = {4},
	pages = {1830-1840},
	doi = {10.1103/PhysRevD.55.1830},
	archivePrefix = {arXiv},
	eprint = {astro-ph/9609170},
	primaryClass = {astro-ph},
	adsurl = {https://ui.adsabs.harvard.edu/abs/1997PhRvD..55.1830Z}
}
\end{document}